\documentclass[fleqn,usenatbib]{mnras}

\usepackage{newtxtext}
\usepackage{mathptmx}
\usepackage{comment}

\usepackage[T1]{fontenc}

\DeclareRobustCommand{\VAN}[3]{#2}
\let\VANthebibliography\thebibliography
\def\thebibliography{\DeclareRobustCommand{\VAN}[3]{##3}\VANthebibliography}

\usepackage{graphicx}	% Including figure files
\usepackage{amssymb}	% Extra maths symbols
\usepackage{amsmath}	% Advanced maths commands

\def\aj{AJ}             	% Astronomical Journal
\def\araa{ARA\&A}       	% Annual Review of Astron and Astrophys
\def\apj{ApJ}           	% Astrophysical Journal
\def\apjl{ApJ}          	% Astrophysical Journal, Letters
\def\apjs{ApJS}         	% Astrophysical Journal, Supplement
\def\aap{A\&A}          	% Astronomy and Astrophysics
\def\aaps{A\&AS}        	% Astronomy and Astrophysics, Supplement
\def\mnras{MNRAS}       	% Monthly Notices of the RAS
\def\pasp{PASP}         	% Publications of the ASP
\def\pasj{PASJ}         	% Publications of the ASJ
\def\nat{Nature}        	% Naturerun_fitpsf3d.pro
\title[SAI analysis for M51, NGC3627 and NGC628]{Instability analysis for spiral arms of local galaxies: M51, NGC3627 and NGC628}

\author[S. Inoue et al.]{
{Shigeki Inoue$^{1,2}$\thanks{E-mail: inouesg@ccs.tsukuba.ac.jp}, Toshinobu Takagi$^{3}$, Atsushi Miyazaki$^{3}$, Erin Mentuch Cooper$^{4}$, Fumi Egusa$^{5}$}
\newauthor{\& Hidenobu Yajima$^{1}$}
\\
$^{1}$Center for Computational Sciences, University of Tsukuba, Ten-nodai, 1-1-1 Tsukuba, Ibaraki 305-8577, Japan\\
$^{2}$Chile Observatory, National Astronomical Observatory of Japan, Mitaka, Tokyo 181-8588, Japan\\
$^{3}$Japan Space Forum, Kanda-surugadai, Chiyoda-ku,Tokyo,101-0062, Japan\\
$^{4}$Department of Astronomy, The University of Texas at Austin, 2515 Speedway, Stop C1400, Austin, Texas 78712, USA\\
$^{5}$Institute of Astronomy, University of Tokyo, Mitaka, Tokyo 181-0015, Japan
}

\date{Accepted XXX. Received YYY; in original form ZZZ}

\pubyear{2020}

\begin{document}
\label{firstpage}
\pagerange{\pageref{firstpage}--\pageref{lastpage}}
\maketitle

% Abstract of the paper, not more than 250 words for Main Journal papers or 200 words for Letters
\begin{abstract}
We investigate dynamical states of grand-design spiral arms in three local galaxies: M51, NGC3627 and NGC628. Based on linear perturbation analysis considering multiple components in the galaxies, we compute instability parameters of the spiral arms using their observational data and argue whether the arms will fragment by their self-gravity. Our analysis utilises observations of carbon monoxide (CO), 21-centimetre line emission and multi-band photometric images for molecular gas, atomic gas and stellar components in the arms, respectively. We find that the grand-design arms of these galaxies indicate marginally stable states, and hence they are not on the way to fragment. We consider this to be consistent with the commonness of spiral galaxies and the relative rarity of fragmented discs at low redshifts. In the analysis, molecular gas is the dominant component to determine the (in)stability of the arms, whereas atomic gas and stars are far less important. Therefore, the results of our analysis are sensitive to an assumed CO-to-H$_{\rm 2}$ conversion factor. If we assume a typical scatter of the measurements and admit nearly twice as large a conversion factor as our fiducial value, our analysis results in predicting the instability for the spiral arms. More sophisticated determination of the conversion factor is required for more accurate analysis for the (in)stability of spiral arms. 
\end{abstract}


%%%%%%%%%%%%%%%%%%%%%%%%%%%%%%%%%%%%%%%%%%%%%%%%%%

%%%%%%%%%%%%%%%%% BODY OF PAPER %%%%%%%%%%%%%%%%%%

\section{Introduction}
\label{Intro}
Spiral arms are common structures of disc galaxies in the local Universe and generally involve active star formation along them. Arms can convey gas from disc regions to the galactic centres due to their non-axisymmetric features, which may drive the formation and growth of (pseudo)bulge and central black holes \citep[e.g.][]{kk:04}. The growth of the bulges and massive black holes increases the central mass concentration of the galaxy and can destruct bar structures \citep[e.g.][]{hn:90,h:12,gdh:20}. In addition, spiral arms are thought to cause radial migration of disc stars by exchanging their angular momenta, which is proposed as a possible formation mechanism of thick discs \citep[e.g.][]{rdq:08,sb:09}. Spiral arms can thus play important roles as drivers of the structural evolution of disc galaxies. However, the physics of generation and evolution of spiral arms has long been under debate \citep[e.g.][]{dobbsbaba:14}.  

The emergence of structured spiral galaxies appears to occur in a relatively recent epoch. \citet{ee:14} showed that the abundance of galaxies with clear spiral structures decreases with redshift, and  they argued that the onset of spiral arms occurs between redshifts $z\simeq1.4$ and $1.8$ from their visual inspection of photometric images. Meanwhile, the most distant spiral galaxies in ever observed have been found at $z=2.54$ \citep{yrg:17}\footnote{The spiral galaxy found by \citet{yrg:17} is magnified by the gravitational lensing effect and cannot be observationally resolved at this redshift without the aid of lensing. They speculated, therefore, that spiral galaxies could exist at even higher redshifts.} and $z=2.18$ \citep{lss:12} with spectroscopic confirmation of their disc rotations. Thus, spiral galaxies become rare with redshift. Instead of spiral arms, a substantial fraction of star-forming galaxies are observed to have giant clumps at redshifts $z\sim2$ \citep[e.g.][]{csh:96,tkt:13II,gfb:14,sok:16}.\footnote{Giant clumps are gas-rich and nearly spherical structures orbiting within a galactic disc, can be as massive as $\sim10^9M_\odot$.} Although clumpy galaxies are also observed at low redshifts, they are quite rare in the local Universe and have significantly higher gas-fractions than typical spiral galaxies \citep[e.g.][]{f:17,fga:17,ofg:18}. The formation of such clumpy structures is often attributed to dynamical instability of their highly gas-rich and therefore dissipative discs \citep[e.g.][]{n:98,n:99}. 

\cite{iy:18} proposed that, in gas-rich discs such as high-redshift galaxies, spiral arms can be unstable by their self-gravity. Such unstable arms are therefore transient structures and can fragment into giant clumps by spiral-arm instability (SAI); the fragmentation of arms is distinct from Toomre instability that characterises fragmentation of a local region in a flat disc \citep{s:60,t:64}. In the low-redshift Universe, since disc galaxies have relatively low gas fractions, their spiral arms are expected to be relatively stable. Accordingly, the clump formation by the spiral-arm fragmentation halts, and such arms are expected to stably exist for a long time. Thus, the scenario proposed by \cite{iy:18} implies that dynamical states of spiral arms may evolve from high to low redshifts. Although it should be noted that the high-redshift clumpy galaxies may not be direct progenitors of local spiral galaxies, the SAI theory possibly explains the evolution of disc galaxies from the clumpy to current states with spirals. 

As the first step to test the above scenario, this paper verifies the stability of spiral arms in local disc galaxies. We apply the SAI analysis of \cite{iy:18} to three galaxies: M51, NGC3627 and NGC628. These galaxies have no giant clumps but two well-defined arms. Such grand-design spiral galaxies are common in the local Universe; \citet{ee:82} have shown that 75 and 57 per cents of barred and non-barred galaxies have two grand-design arms in their sample. This kind of arms is generally thought to form by large-scale processes such as interactions with their companion galaxies \citep[e.g.][]{bt:08}. Therefore, the time-scales of their presence would be comparable to those of the interactions, i.e. $t\lesssim1~{\rm Gyr}$ \citep[e.g.][]{dtp:10}, and it is expected that their formation took place hundreds of million years ago. However, there is no guarantee that the grand-design arms are stable and not on the way to fragmentation. The SAI analysis is able to predict whether spiral arms fragment or not \citep{iy:18}. For our sample galaxies, we calculate instability parameters of their spiral arms using archived data of radio observations for line emission of carbon monoxide (CO) tracing molecular hydrogen (H$_{\rm 2}$), the 21-centimetre line from atomic hydrogen (H$_{\rm I}$) and multi-band photometric observations to obtain stellar surface density. Then, using the SAI analysis, we address the dynamical states of the spiral arms in the local galaxies.  

We organise this paper as follows. In Section \ref{ana}, we describe the basic equations of the SAI analysis and the computation of the instability parameter (more details in Appendix \ref{lpa}). We explain the observational data of our sample galaxies in Section \ref{obs} and our methods to obtain physical quantities required for the instability analysis in Section \ref{utilize}. Section \ref{res} shows our results, and Section \ref{dis} discusses the dynamical states of the galaxies according to the results. Section \ref{con} presents our conclusions and a summary of this study.

\section{Instability analysis}
\label{ana}
The SAI analysis of \cite{iy:18} is based on the local linear perturbation theory of \citet{tti:16}. It considers azimuthal perturbations propagating along a spiral arm that is assumed to have a negligible pitch angle, i.e. the tight-winding approximation. We present the details of the linear perturbation analysis in Appendix \ref{lpa} \citep[see also][]{tti:16,iy:19,iy:20,iyh:21}.

A spiral galaxy generally consists of various components such as cold and warm gas and stars with different ages, and their dynamical states can be different from each other. Although we apply our analysis to such a multi-component spiral arm, we begin with considering the case of a single-component arm. For the component $i$ of an arm, its line-mass (mass of arm per unit length), azimuthal velocity dispersion, half-width and vertical thickness are denoted by $\Upsilon_i$, $\sigma_i$, $W_i$ and $h_i$, respectively. For an azimuthal perturbation of wavenumber $k$, the instability parameter for the component $i$ is given as 
\begin{equation}
S_i(k)=\frac{\sigma_i^2k^2 + \kappa_i^2}{\upi G\Upsilon_i f(kW_i)F(kh_i) k^2},
    \label{S_each}
\end{equation}
where $\kappa_i$ is epicyclic frequency and $f(kW_i)\equiv[K_0(kW_i)L_{-1}(kW_i) + K_1(kW_i)L_0(kW_i)]$;\footnote{The function $f(kW_i)$ describes the shape of a perturbed Poisson equation for a Gaussian ring (see Appendix \ref{AppSingle}).} here $K_j$ and $L_j$ are modified Bessel and Struve functions of order $j$. The function $F(kh_i)=[1-\exp(-kh_i)]/(kh_i)$ gives a thickness correction factor \citep[see][]{t:64}. On the right-hand side, the denominator represents the force inwards by self-gravity, and the numerator does the force outwards by internal pressure and the Coriolis force. When $S_i(k)=1$, the forces inwards and outwards balance. Accordingly, when $S_i(k)<1$, the perturbation $k$ is expected to grow rapidly by the self-gravity and be dynamically unstable.

In this study, we assume a spiral arm to consist of three components: molecular gas (H$_{\rm 2}$), atomic gas (H$_{\rm I}$) and stars. We consider that these components interact with each other through gravity, and their dynamical states are connected via the Poisson equation (see Appendix \ref{AppMulti}). Then, the total instability parameter of the arm is given as 
\begin{equation}
    S_{\rm tot}(k) = \left[\frac{1}{S_{\rm H_2}(k)} + \frac{1}{S_{\rm H_I}(k)} + \frac{1}{S_{\rm s}(k)}\right]^{-1}.
    \label{S_total}
\end{equation}
Again, the perturbation $k$ is expected to be unstable when $S_{\rm tot}(k)<1$, and vice versa. Note that $S_{\rm tot}$ is a function of $k$. Hence, the instability condition of the arm is $\min[S_{\rm tot}(k)]<1$.

Equation (\ref{S_total}) implies that $S_{\rm tot}$ is always lower than $S_i$ of any component. Therefore, if there is a component with $S_i<1$, the total instability parameter is necessarily $S_{\rm tot}<1$ and predicts the instability for the arm. However, this is not vice versa; even if all components have $S_i>1$, their combined value can be $S_{\rm tot}<1$. We can compute the most unstable wavelength as $\lambda_{\rm MU}=2\pi/k_{\rm MU}$, where $k_{\rm MU}$ gives the minimum value of $S_{\rm tot}$: $S_{\rm tot}(k_{\rm MU})=\min[S_{\rm tot}(k)]$. When $\min(S_{\rm tot})<1$, $\lambda_{\rm MU}$ can be taken as a physical scale of a collapsing segment of the unstable arm \citep{iy:18}. 

We again emphasise that the SAI predicted by equation (\ref{S_total}) is physically distinct from Toomre instability. Because the Toomre analysis considers a local region in a flat disc, it does not have the parameter of arm width $W_i$. Moreover, since a spiral arm is considered to be a non-linear structure in a disc, the (in)stability of the arm is essentially beyond the applicable domain of the Toomre analysis. Because the Toomre instability operates in a two-dimensional region, it is ambiguous what structure the instability finally forms, a spiral arm, ring or clump. In the SAI analysis, if $W_i$ is significantly smaller than $\lambda_{\rm MU}$, the instability is expected to one-dimensionally collapse along the spiral arm and form a clump.\footnote{When an arm is significantly wider than the perturbation wavelength with $kW_i\gg1$, the SAI analysis reduces to Toomre analysis for azimuthal perturbations (see Appendix \ref{AppSingle}).}

\citet{iy:18} performed simulations of disc galaxies in isolation and applied the SAI analysis to their simulation results. Although their simulated discs consisted of two components (gas and stars) in rigid potentials of dark matter haloes, they demonstrated that the SAI analysis can predict with accuracy whether or not a spiral arm fragments and estimate a mass of the clump formed by the fragmentation. \cite{iy:20} developed the SAI analysis to include the thickness correction $F(kh)$ and applied it to cosmological simulations of circum-stellar discs of first stars, where a spiral arm fragments into secondary stars \citep[e.g.][]{gbc:12}. They showed that the analysis can characterise the fragmentation of arms in such three-dimensional simulations including gas cooling. These previous studies utilised their simulation data to compute $S_{\rm tot}$. In this paper, we apply the SAI analysis to observational data basically in the same way as in these studies (see Section \ref{utilize}).

\section{Data of the sample galaxies}
\label{obs}
Computing equations (\ref{S_each} and \ref{S_total}) requires obtaining the quantities of $\Upsilon_i$, $\sigma_i$, $\kappa_i$, $W_i$ and $h_i$ for each component. In order to perform local analysis using $S_{\rm tot}$, we use archived data and modellings described below. 

To obtain two-dimensional distributions of surface densities and velocity dispersions of H$_{\rm 2}$ and H$_{\rm I}$ components, we use radio observations for rotational transition lines of $^{12}$C$^{16}$O and 21-centimetre line, respectively. For stellar components, we use multi-band photometric observations and perform spectral energy distribution (SED) fitting to estimate stellar surface densities of local regions. For rotation velocities, we use rotation curves measured by previous studies.

\subsection{M51}
\label{m51}
M51, also known as the Whirlpool galaxy and NGC5194, is a nearby grand-design spiral galaxy. It is generally thought that the formation of the arms is driven by the close interaction with its companion NGC5195. \citet{mwf:12} performed spatially resolved SED fitting on M51 and provide us with a two-dimensional map of mass density for the galaxy. They estimated the total stellar and gas masses to be $4.7$ and $1.1\times10^{10}~{\rm M_\odot}$, respectively (excluding the companion).

For the molecular component, we use the publicly available data of PAWS\footnote{https://www2.mpia-hd.mpg.de/PAWS/PAWS/Home.html} \citep[the PdBI Arcsecond Whirlpool Survey,][]{PAWS} for zeroth- and second-moment maps of CO(1-0) lines observed by PdBI and the IRAM 30m telescope with a resolution of $\sim1~{\rm arcsec}$ in the inner region of $R\lesssim5~{\rm kpc}$. The moment maps are created by cumulating pixels in two consecutive velocity channels that have signals above $4\sigma_{\rm rms}$, and these are extended to include adjacent pixels that have signals above $1\sigma_{\rm rms}$ in at least two consecutive channels \citep{chs:14}. Molecular surface densities $\Sigma_{\rm H_2}$ are estimated from velocity-integrated surface brightness temperatures $T_{\rm CO(1-0)}$ as $\Sigma_{\rm H_2}=\alpha_{\rm CO(1-0)}T_{\rm CO(1-0)}$, where we assume the CO-to-H$_{\rm 2}$ conversion factor $\alpha_{\rm CO(1-0)}=3.5~{\rm M_\odot~pc^{-2}~(K~km~s^{-1})^{-1}}$ including the contribution of helium. This value of $\alpha_{\rm CO(1-0)}$ is measured for selected regions in the spiral arms of M51 \citep{swa:10}.\footnote{\citet{swa:10} estimated that $\alpha_{\rm CO(1-0)}$ is from $2.8$ to $4.3~{\rm M_\odot~pc^{-2}~(K~km~s^{-1})^{-1}}$. We adopt is the central value of this range.} We argue uncertainties of $\alpha_{\rm CO(1-0)}$ in Section \ref{alphaCO}. Although there are several methods to estimate a velocity dispersion of gas from a line profile, we use the second moment of flux densities as a velocity dispersion.

For the H$_{\rm I}$ component, we use the public data of THINGS\footnote{https://www2.mpia-hd.mpg.de/THINGS/Overview.html} \citep[The H$_{\rm I}$ Nearby Galaxy Survey,][]{THINGS}. The observed maps we use are created with the {\sc Robust} weighting scheme \citep{briggs:95}, and the major and minor axes of the synthesized beam are $5.82$ and $5.56~{\rm arcsec}$ for M51. Similarly to the case of H$_2$, atomic surface densities $\Sigma_{\rm H_I}$ are estimated as $\Sigma_{\rm H_I}=\alpha_{\rm 21cm}T_{\rm 21cm}$, where $T_{\rm 21cm}$ is velocity-integrated surface brightness temperatures of 21-centimetre line, and the conversion factor is $\alpha_{\rm 21cm}=1.98\times10^{-2}~{\rm M_\odot~pc^{-2}~(K~km~s^{-1})^{-1}}$ including the helium contribution \citep{THINGS}. We consider the second moments of the 21-centimetre line to be the velocity dispersions $\sigma_{\rm HI}$ of the atomic gas.

For the stellar component, we use the result of \citet{mwf:12} where their stellar SED fitting is performed with multi-wavelength analysis including 7 photometric images from optical to near-infrared wavelengths: $B$-, $V$-, $R$- and $I$-bands of SINGS \citep[Spitzer Infrared Nearby Galaxies Survey,][]{SINGS} and $J$-, $H$- and $K$-bands of the 2MASS\footnote{https://irsa.ipac.caltech.edu/Missions/2mass.html} (Two Micron All Sky Survey) Large Galaxy Atlas \citep{jcc:03}. The images are remapped to a common spatial resolution of $28~{\rm arcsec}$ in full-width at half maximum (FWHM) and plate scale of $10~{\rm arcsec}$ per pixel. The initial mass function (IMF) of \citet{KroupaIMF} is adopted. Since we cannot obtain velocity dispersions of stars from the data, we model the two-dimensional distribution of $\sigma_{\rm s}$ (see Section \ref{disp}).

\begin{figure}
  \includegraphics[bb=0 0 736 770, width=\hsize]{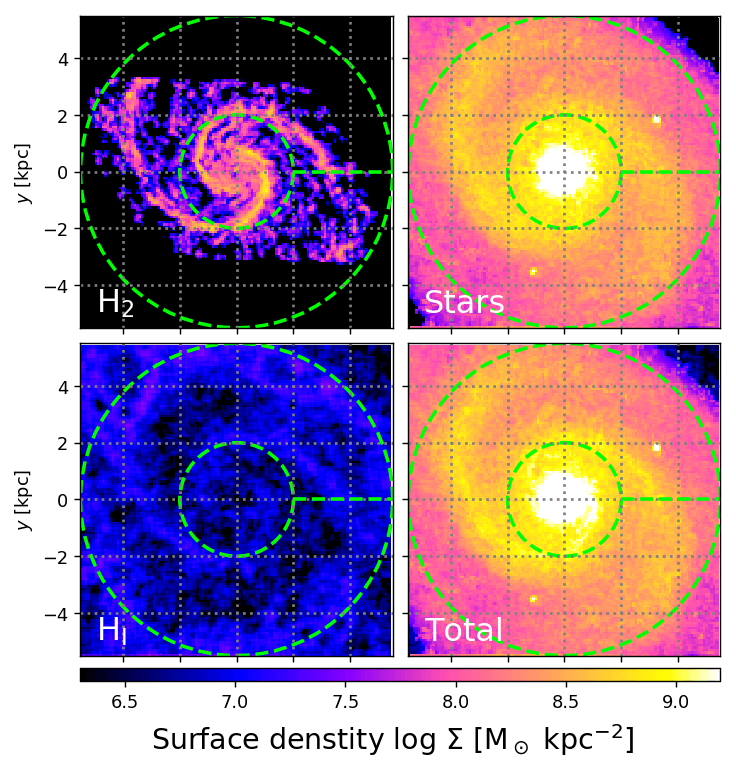}
  \caption{Surface density distributions of molecular (top left), atomic gas (bottom left) and stars (top right) for M51. The summation of the three densities is shown in the bottom right panel. All maps are deprojected on to the disc plane at the distance $d=7.6~{\rm Mpc}$. Note that the CO(1-0) observations of PAWS only cover the parallelogram region in the top left panel. We apply the SAI analysis to the region between the inner and outer circles delineated by the green dashed lines. The horizontal green dashed line along $y=0$ corresponds to the origin of angular coordinate $\phi=0$ in our polar-map analysis shown in Section \ref{m51res}; $\phi$ increases anticlockwise.}
  \label{M51_Maps}
\end{figure}
We assume the distance to M51 to be $d=7.6~{\rm Mpc}$ \citep{cfj:02,PAWS}, and the inclination and position angle are $i_{\rm inc}=27^\circ$ and $\phi_{\rm PA}=162^\circ$ \citep{os:14}. Fig. \ref{M51_Maps} shows the face-on surface density maps deprojected on to the disc plane at the assumed distance. It should be noted that the PAWS observations only cover the inner region of the galaxy (the top left panel). We therefore restrict our analysis to the region inside the outer circles (the green dashed lines) in Fig. \ref{M51_Maps}, and this study argues the (in)stability of the `inner arms' of M51. The surface densities of H$_{\rm I}$ are significantly lower than those of the other components, and most of the gas in the disc of M51 appears to be in the molecular form \citep{kss:09}. This implies that the H$_{\rm I}$ component is expected to be less important than H$_{\rm 2}$ and stars in the SAI analysis (see Section \ref{m51res}).

For M51, \citet{os:14} have measured rotation velocities of gas to be $v_\phi\simeq200~{\rm km~s^{-1}}$ at all radii inside $R\lesssim7.5 ~{\rm kpc}$.\footnote{They argued that the gas disc of M51 suddenly bends at $R\sim7.5 ~{\rm kpc}$, and this can be taken as the impact of the interaction with its companion.} Because this radius encircles the inner arms observed by PAWS, we adopt this constant velocity. We assume that all the components of H$_{\rm 2}$, H$_{\rm I}$ and stars share the same rotation velocity and disc plane.

\subsection{NGC3627}
\label{ngc3627}
NGC3627, also known as M66, is a member of the Leo Triplet. Its two grand-design arms are somewhat asymmetric, which may be indicative of disturbance by other group members. This galaxy has a bar structure, and the roots of the spiral arms are contiguous with the tips of the bar. In previous studies, the total stellar, H$_{\rm I}$ and H$_2$ masses have been estimated to be $4.0\times10^{10}$, $8.2\times10^8$ and $2.8\times10^9~{\rm M_\odot}$, respectively \citep{lwb:08,THINGS,sls:18}.

For the H$_{\rm 2}$ component, we utilise archived data obtained by the PHANGS-ALMA project\footnote{https://sites.google.com/view/phangs/home} \citep[Physics at High Angular resolution in Nearby GalaxieS using the Atacama Large Millimeter/Submillimeter Array, e.g.][]{PHANGS-ALMA}. The observations are performed by the ALMA programme 2015.1.00956.S (P.I. A. K. Leroy), and we combine their data for CO(2-1) line emission with band 6, which were obtained by the 12-metre array, the Morita Atacama Compact Array and the total power observations. Because the original data are separated into northern and southern blocks of the galaxy, we perform self-calibration and imaging processes for each block using {\sc CASA} (Common Astronomy Software Applications) and combine them into a single mosaic cube image. We compute the zeroth- and second-moment maps of CO(2-1) emission of the whole region including the two spiral arms. The major and minor axes of the resultant synthesized beam are $1.15$ and $1.00~{\rm arcsec}$. The values of rms are $\sigma_{\rm rms}=7.2$ and $4.8~{\rm mJy}$ per beam in the northern and southern sides of this galaxy. The velocity resolution is $2.5~{\rm km~s^{-1}}$. The moment maps are created with the same masking technique as was done in PAWS for M51 (see Section \ref{m51}). Surface densities of H$_{\rm 2}$ are estimated from the zeroth-moment map of CO(2-1) emission as $\Sigma_{\rm H_2}=\alpha_{\rm CO(1-0)}T_{\rm CO(2-1)}/r_{\rm 21}$, where we assume the CO(2-1)-to-CO(1-0) line ratio to be $r_{\rm 21}\equiv T_{\rm CO(2-1)}/T_{\rm CO(1-0)}=0.79$ \citep{xclodgass:17} and again $\alpha_{\rm CO(1-0)}=3.5~{\rm M_\odot~pc^{-2}~(K~km~s^{-1})^{-1}}$ measured in the arms of M51.

For the H$_{\rm I}$ component, we use the THINGS data. The major and minor axes of the synthesized beam are $5.67$ and $5.45~{\rm arcsec}$ for this galaxy \citep{THINGS}. Atomic surface densities are estimated via the same conversion factor $\alpha_{\rm 21cm}$ as used for M51 (Section \ref{m51}).

For the stellar component, we use the SED fitting code {\sc Bagpipes}\footnote{https://bagpipes.readthedocs.io/en/latest/} \citep[Bayesian Analysis of Galaxies for Physical Inference and Parameter EStimation, ][]{BAGPIPES} to estimate stellar surface densities. We input 10 photometric images: far- and near-ultraviolet (UV) bands of GALEX\footnote{https://archive.stsci.edu/missions-and-data/galex-1/} (the Galaxy Evolution Explorer), $u$-, $g$-, $r$-, $i$- and $z$-bands of SDSS\footnote{https://www.sdss.org/} (the Sloan Digital Sky Survey) and $J$-, $H$- and $K$-bands of 2MASS. The SED fitting for this galaxy uses the UV bands to constrain mass of young stars, whereas that for M51 uses the optical bands. The fitting using the UV images is more sensitive to young stars. These images are remapped to a spatial resolution of $\sim6~{\rm arcsec}$ in FWHM and a plate scale of $1.5~{\rm arcsec}$ per pixel while matching their point spread functions to that of the near-UV image. In the SED fitting, the IMF of \citet{kb:02} is adopted. We employ a two-population model in which the population older than $1~{\rm Gyr}$ have star formation rates decaying exponentially with time, whereas the other population forms instantaneously. This is because the fitting is often too strongly biased to young stars in UV-bright regions. We adopt the extinction curves of \citet{cab:00} for dust attenuation. For each pixel, {\sc Bagpipes} searches for the best-fitting parameters by matching the modelled SED with the observed flux densities weighted by their signal-to-noise ratios (SNRs). Here, in order to avoid too strong bias to observations with high SNRs, we impose an upper limit of SNR$=50$. We take into account pixels that have SNR$>3$ in at least 6 bands.

\begin{figure}
  \includegraphics[bb=0 0 736 770, width=\hsize]{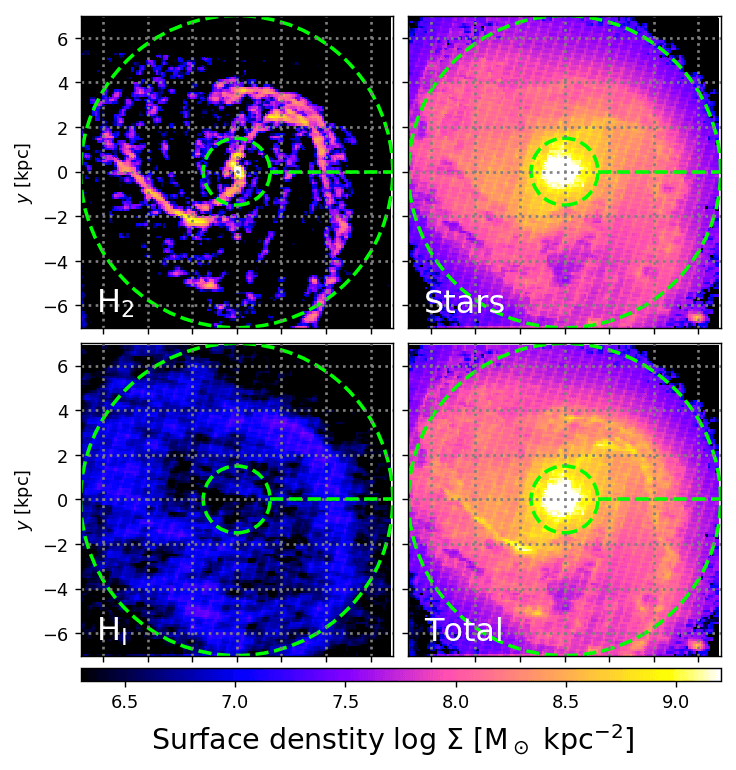}
  \caption{Same as Fig. \ref{M51_Maps} but for NGC3627.}
  \label{NGC3627_Maps}
\end{figure}
For NGC3627, we assume its distance, inclination and position angle to be $d=10.6~{\rm Mpc}$, $i_{\rm inc}=57.3^\circ$ and $\phi_{\rm PA}=173.1^\circ$ \citep{PHANGS-rotation}. Fig. \ref{NGC3627_Maps} shows the deprojected face-on maps of the surface densities. As in the case of M51, the H$_{\rm I}$ gas also is quite diffuse in this galaxy. Although stellar arms are not clearly seen in the top right panel, the molecular arms outshine in the top left and bottom right panels. Artificial stripes appear in the deprojected stellar map due to the relatively high $i_{\rm inc}$. However, this does not affect our SAI analysis because of the smoothing applied in Section \ref{polarmap}.

\citet{PHANGS-rotation} has performed detailed kinematic analysis and obtained rotation curves using CO lines for the PHANGS-ALMA sample. The observed rotation curves are fitted with a function,
\begin{equation}
    v_\phi=v_0\frac{2}{\upi}\arctan\left(\frac{R}{R_{\rm t}}\right),
    \label{tancurve}
\end{equation}
and they found $v_0=202.1~{\rm km~s^{-1}}$ and $R_{\rm t}=0.4~{\rm kpc}$ for NGC3627. We use this function and the parameters for all the components in our analysis.

\subsection{NGC628}
\label{ngc628}
NGC628, also known as M74, is a face-on galaxy without clear bar structures. Because it belongs to a group of galaxies, the formation of its grand-design arms may be driven by interactions with other group members. The total stellar, H$_{\rm I}$ and H$_2$ masses have been estimated to be $1.3\times10^{10}$, $3.8\times10^9$ and $6.7\times10^8~{\rm M_\odot}$\footnote{This H$_2$ mass is determined by measurements at 45-pc resolution; however the mass increases to $1.0\times10^9~{\rm M_\odot}$ at 120-pc resolution \citep{sls:18}.} in previous studies \citep{lwb:08,THINGS,sls:18}.

For the H$_{\rm 2}$ component, we use the data presented in \citet{hph:20}, in which they obtained CO(1-0) line emission of the galaxy by the ALMA programme 2013.1.00532.S (P.I. E. Schinnerer) and made the data cube publicly available.\footnote{https:\slash\slash{}www.canfar.net\slash{}storage\slash{}list\slash{}AstroDataCitationDOI\slash{}CISTI.CANFAR\slash{}20.0001\slash{}data} The velocity resolution is $6.0~{\rm km~s^{-1}}$, and the image rms is $\sigma_{\rm rms}=0.12~{\rm K}$. In creating the moment maps, we adopt the same method of the masking criterion as in the cases of M51 and NGC3627 (see Section \ref{m51}). Surface densities of H$_{\rm 2}$ are estimated via the same CO(1-0)-to-H$_{\rm 2}$ conversion factor $\alpha_{\rm CO(1-0)}=3.5~{\rm M_\odot~pc^{-2}~(K~km~s^{-1})^{-1}}$.

The H$_{\rm I}$ observations are taken from the THINGS data. The major and minor axes of the synthesized beam are $6.8$ and $5.57~{\rm arcsec}$ \citep{THINGS}. The conversion factor $\alpha_{\rm 21cm}$ is the same as used for M51 and NGC3627.

The stellar surface densities are estimated with the same method as in NGC3627 (see Section \ref{ngc3627}). The images are remapped to the resolution of $\sim5~{\rm arcsec}$ in FWHM and the plate scale of $1.5~{\rm arcsec}$ per pixel. Because of lower SNRs in NGC628, we take into account pixels that have SNR$>3$ in at least 5 bands. Foreground stars brighter than the $J$-band magnitude of $16$ are masked with a radius of $6~{\rm arcsec}$.

\begin{figure}
  \includegraphics[bb=0 0 736 770, width=\hsize]{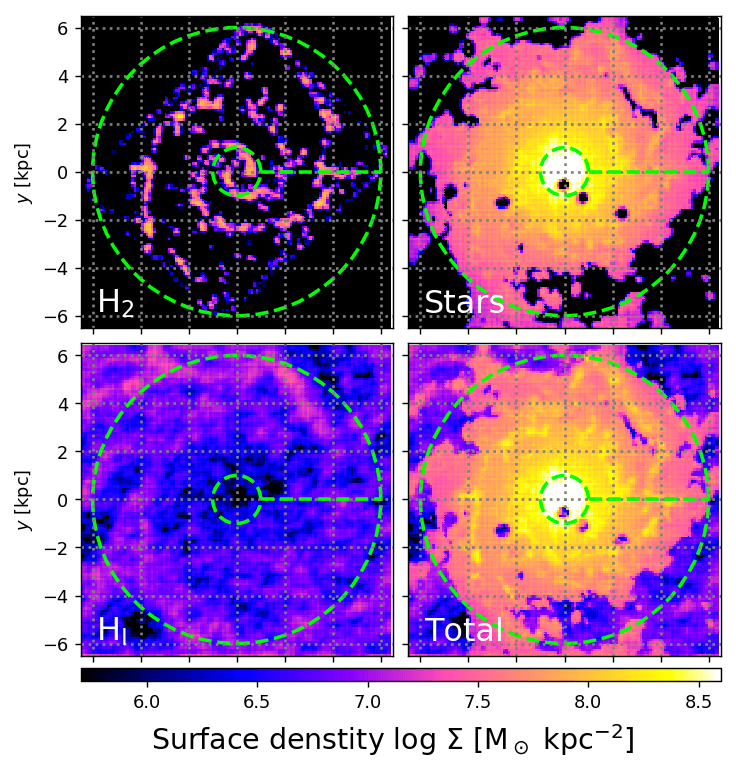}
  \caption{Same as Figs. \ref{M51_Maps} and \ref{NGC3627_Maps} but for NGC628. Note that the colour scales to indicate the densities are different from those in Figs. \ref{M51_Maps} and \ref{NGC3627_Maps}; NGC628 appears to have relatively lower densities than M51 and NGC3627.}
  \label{NGC628_Maps}
\end{figure}
For NGC628, we assume its distance, inclination and position angle to be $d=9.8~{\rm Mpc}$, $i_{\rm inc}=8.9^\circ$ and $\phi_{\rm PA}=20.7^\circ$ \citep{PHANGS-rotation}. Fig. \ref{NGC628_Maps} shows the deprojected surface densities. As seen in the H$_{\rm 2}$ density map (top left panel), the CO observations only cover a nearly rectangular region of $\sim8~{\rm kpc}$ on a side. Therefore, we restrict our analysis to the region inside $R=6~{\rm kpc}$ from the galactic centre (the outer green dashed circles in Fig. \ref{NGC628_Maps}). Although the spiral arms are diffuse in the H$_{\rm I}$ and stellar maps, the two symmetric arms are seen in the H$_{\rm 2}$ distribution. The total surface densities of this galaxy appear to be relatively lower than those of M51 and NGC3627.

The rotation velocites of this galaxy are also fitted with equation (\ref{tancurve}), and \citet{PHANGS-rotation} derived the parameters to be $v_0=144.8~{\rm km~s^{-1}}$ and $R_{\rm t}=0.6~{\rm kpc}$ for NGC628.

\section{Derivation of the physical quantities}
\label{utilize}
\subsection{Coordinate conversion}
\label{polarmap}
It is useful for the SAI analysis to transform the deprojected zeroth- and second-moment maps into the polar coordinates $(R,\phi)$. In this procedure, we apply a two-dimensional Gaussian kernel to the zeroth-moment maps and make the pixelation of the three components consistent. The kernel has a standard deviation of $w_{\rm ker}=0.21~{\rm kpc}$. We also process the second-moment maps in the same way while weighting by the surface densities.

\subsection{Line-mass and half-width}
\label{armfit}
Line-mass $\Upsilon_i$ and half-width $W_i$ of the arms are computed from a two-dimensional map of surface densities $\Sigma_i$. In the polar coordinates, we perform one-dimensional Gaussian fitting along the radial direction at a given $\phi$. The fitting function is defined as $\tilde{\Sigma_i}(R,\xi,\phi)=\Sigma_i(R,\phi)\exp(-\xi^2/2w^2)$, where $\xi$ represents radial offset from $R$, and $\Sigma_i(R,\phi)$ is a surface density obtained in Section \ref{polarmap}. The fitting is iteratively applied in the range of $-1.55w<\xi<1.55w$ while changing $w$ for each component, and we obtain the best-fitting value of $w$ that gives the minimum value of goodness-of-fit $\chi_i^2$. If there is a crest of the arm at $R$ and the radial density distribution is nearly Gaussian, $\chi_i^2$ becomes significantly lower than unity. We define the edges of the arm as the inner and outer radii where $\tilde{\Sigma}(R,\pm W,\phi)=0.3\Sigma(R,\phi)$ in the fitting function, and this definition gives the half-width to be $W_i=1.55w$. The line-mass of the arm is computed as 
\begin{equation}
\Upsilon_i(R,\phi)=\int^{W_i}_{-W_i}\tilde{\Sigma_i}(R,\xi,\phi)~\textrm{d}\xi=AW_i\Sigma_i(R,\phi),
\label{linemass}
\end{equation}
where $A=1.44$ for a Gaussian distribution \citep[][]{tti:16}. 

Because of the Gaussian kernel applied in the coordinate conversion (Section \ref{polarmap}), $W_i$ calculated above is overestimated although $\Upsilon_i$ is little affected. By subtracting the broadening effect by the kernel, the intrinsic half-width is approximated as $W'_i=[W_i^2 - (1.55w_{\rm ker})^2]^{1/2}$. We use this intrinsic half-width as $W_i$ in equation (\ref{S_each}) although this alteration does not change $S_i(k)$ significantly. 

\subsection{Epicyclic frequency}
\label{kapa}
Although our analysis can give each component its own epicyclic frequency $\kappa_i$ (equation \ref{S_each}), we assume that all of the three components have the same rotation curve, i.e. $\kappa_{\rm H_2}=\kappa_{\rm H_I}=\kappa_{\rm s}$. We confirmed that the first-moment maps of CO and 21-centimetre lines are not significantly different from each other in our sample galaxies. Furthermore, we ignore its azimuthal variation and compute $\kappa$ as a function of $R$ from the observed rotation curves. Since self-gravitating arms are expected to have rigid rotations, we assume $\kappa=2\Omega=2v_\phi/R$ \citep[see][]{tti:16,iy:18}. We discuss the validity of this assumption in Section \ref{others}.

\subsection{Azimuthal and vertical velocity dispersions}
\label{disp}
We consider the second moments of CO and 21-centimetre lines to be velocity dispersions of H$_{\rm 2}$ and H$_{\rm I}$ gas. Note that these are dispersions of turbulent velocities and should include the contributions by thermal and magnetic pressure in the dynamical analysis. It is, however, challenging to measure temperatures and magnetic fields of the molecular and atomic gas in the spiral arms. In this study, we ignore the thermal and magnetic pressure; however we discuss the significance of these effects in Section \ref{others}. We consider that the velocity dispersions of the gas components are isotropic; namely azimuthal dispersions $\sigma_i$ are the same as vertical ones $\sigma_{z,i}$ for the H$_{\rm 2}$ and H$_{\rm I}$ components.

It is quite difficult to observationally determine velocity dispersions of stars in gas-rich regions such as spiral arms. Hence, we employ a disc model of \citet{lwb:08}, where a local value of vertical velocity dispersion of stars is approximated as 
\begin{equation}
    \sigma_{z,{\rm s}}^2(R,\phi)=\frac{2\upi Gl_{\rm s}}{q_{\rm s}}\Sigma_{\rm s}(R,\phi),
    \label{L08model}
\end{equation}
where $l_{\rm s}$ is scale-length of stellar disc, and $q_{\rm s}$ is ratio between scale length and height. We adopt $l_{\rm s}=2.7$, $3.2$ and $3.1~{\rm kpc}$ for M51, NGC3627 and NGC628\footnote{These values of $l_{\rm s}$ are based on those calculated by \citet{lwb:08} with their exponential disc fitting but modified according to the galactic distances assumed in this study.} \citep{lwb:08}, the disc aspect ratio $q_{\rm s}=7.3$ for the three galaxies, which is the averaged value measured in \citet{kkg:02}. Stellar velocity distribution is not necessarily isotropic because of the collisionless nature \citep{bt:08}, and we therefore assume a fixed ratio between vertical and radial dispersions $\sigma_{z,{\rm s}}/\sigma_{R,{\rm s}}=0.6$ \citep{sgm:03}. Finally, using the epicyclic approximation \citep{bt:08}, we compute azimuthal dispersion as
\begin{equation}
    \sigma_{\rm s}^2(R,\phi) = \frac{\kappa'^2}{4\Omega^2}\sigma^2_{R,{\rm s}}(R,\phi),
    \label{sigmaphistars}
\end{equation}
where
\begin{equation}
  \kappa'^2=2\frac{v_{\phi}}{R}\left(\frac{\mathrm{d}v_{\phi}}{\mathrm{d}R} + \frac{v_{\phi}}{R}\right),
  \label{kappa}
\end{equation} 
where we use the observed rotation curves for $v_{\phi}$ and ignore its azimuthal variation. Note that $\kappa'$ in the above equations is different from $\kappa$ in equation (\ref{S_each}). The latter $\kappa$ is a value in a self-gravitating spiral arm assumed to have a rigid rotation with $\kappa=2\Omega$ (see Section \ref{kapa}), whereas the former $\kappa'$ is for the disc model. It is unclear whether the above model of \citet{lwb:08} considering a flat disc can be applied to a spiral arm, and the model can involve various uncertainties and inaccuracies. However, stars are not determinant components of the (in)stability in our analysis. Therefore, our results do not significantly depend on the details of the disc model to estimate $\sigma_{\rm s}$ and $\sigma_{z,{\rm s}}$  (see Sections \ref{res} and \ref{others}). 

\subsection{Vertical thickness}
\label{thickness}
The thickness correction factor $F(kh_i)$ in equation (\ref{S_each}) represents the effect that makes a gravitational potential shallower by lowering mass concentration on the disc plane due to thickness. For each component, we define a vertical thickness of a spiral arm as $h_i\equiv\sigma_{z,i}^2/(\pi G\Sigma_i)$, where $\sigma_{z,i}$ is vertical velocity dispersion of the $i$-th component computed in Section \ref{disp}. Although previous studies propose different correction factors \citep{gl:65,v:70,r:92,rw:11}, the differences between them are not significant \citep[see also][]{e:11}. Since we have two-dimensional distributions of $\Sigma_i$ and $\sigma_{z,i}$, the values of $h_i$ are computed in the polar coordinates $(R,\phi)$ for each component.

\section{Results}
\label{res}
To summarise the data analysis in Section \ref{utilize}, we obtain $\Sigma_i$, $\sigma_i$, $W_i$ and $h_i$ as functions of $(R,\phi)$ for each component, and the unified $\kappa$ as a function of $R$. In what follows, we compute equations (\ref{S_each} and \ref{S_total}) and obtain the local instability parameter $\min(S_{\rm tot})$ on the polar coordinates for M51, NGC3627 and NGC628.

\subsection{M51}
\label{m51res}
\begin{figure*}
  \includegraphics[bb=0 0 2066 1087, width=\hsize]{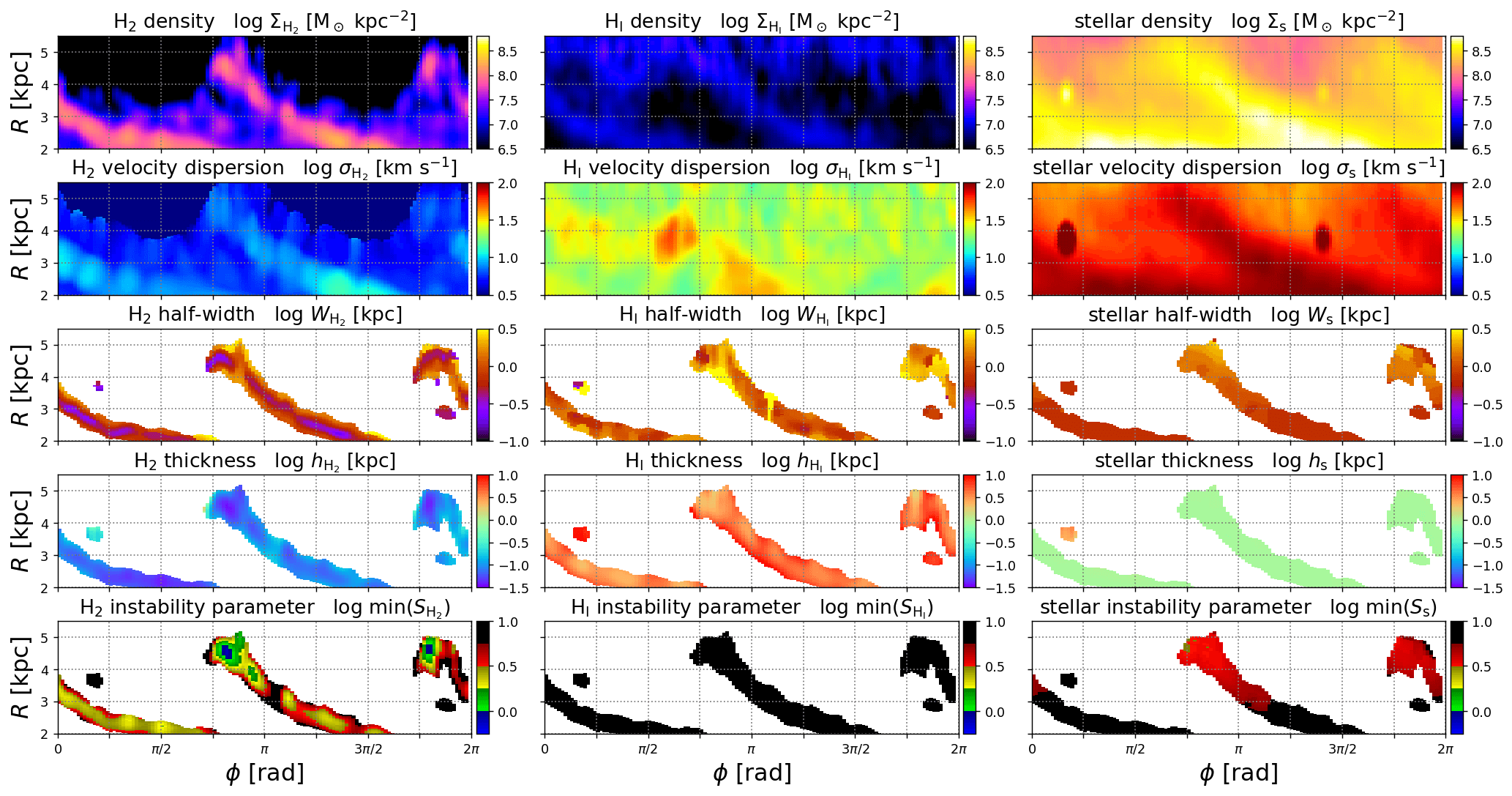}
  \caption{Our polar-map analysis for M51: surface densities ($\Sigma_i$), azimuthal velocity dispersions ($\sigma_i$), half-widths of arm ($W_i$), vertical thicknesses ($h_i$) and the single-component instability parameters [$\min(S_i)$], from top to bottom panels. The left, centre and right columns indicate the results for the H$_{\rm 2}$, H$_{\rm I}$ and stellar components, respectively. The colour scales are logarithmic in all panels. The origin of the angular coordinate, $\phi=0$, corresponds to the horizontal green dashed line in Fig. \ref{M51_Maps}. In the three rows from the bottom, the inter-arm regions where $\log[(\chi^2_{\rm H_2}+\chi^2_{\rm H_I}+\chi^2_{\rm s})/3]>-0.25$ are uncoloured, according to our definition of the spiral-arm regions. In the bottommost panels, blue colours correspond to unstable regions with $\min(S_i)<1$ for each component.}
  \label{M51_Ingre}
\end{figure*}
Fig. \ref{M51_Ingre} shows our polar-map analysis for $\Sigma_i$, $\sigma_i$, $W_i$, $h_i$ and $\min(S_i)$ for the H$_{\rm 2}$, H$_{\rm I}$ and stellar components of M51. In the polar maps, the origin of the angular coordinate $\phi=0$ corresponds to the horizontal green dashed line in Fig. \ref{M51_Maps}, and $\phi$ increases anticlockwise. We define the spiral arms of M51 to be the regions where $\log[(\chi^2_{\rm H_2}+\chi^2_{\rm H_I}+\chi^2_{\rm s})/3]<-0.25$ in the radial Gaussian fittings (Section \ref{armfit}). Although this threshold of $\chi_i^2$ is arbitrary, the computations of $S_{\rm tot}$ are independent of it. In the three bottom rows of Fig. \ref{M51_Ingre}, the areas inside and outside the spiral-arm regions are coloured and uncoloured. The spiral-arm regions detected by the above threshold are consistent with the high-density regions in the topmost panels for $\Sigma_i$. Thus, the two spiral arms are correctly captured with our arm-detection scheme.

In the panels for $\sigma_i$ (the second rows), although the H$_{\rm I}$ velocity dispersions do not vary significantly between inside and outside the arms, the H$_{\rm 2}$ velocity dispersions increase inside the arms. Although the stellar velocity dispersions also increase inside the arms, this is because of the high $\Sigma_{\rm s}$ via equation (\ref{L08model}). 

The panels in the third and fourth rows in Fig. \ref{M51_Ingre} indicate half-widths $W_i$ and vertical thicknesses $h_i$ of the arms. As expected from Fig. \ref{M51_Maps}, the H$_{\rm 2}$ arms indicate the lowest widths in the three components: $W_{\rm H_2}\lesssim0.5~{\rm kpc}$. The stellar arms have $W_{\rm s}\sim1~{\rm kpc}$, and the H$_{\rm I}$ arms appear to be wider than the other components. Although it is noted that their vertical thicknesses $h_i$ are not measured directly but estimated from their velocity dispersions and surface densities by assuming the equilibrium (see Section \ref{thickness}), the H$_{\rm 2}$ arms are vertically thin: $h_{\rm H_2}\lesssim0.1~{\rm kpc}$. On the other hand, the H$_{\rm I}$ arms are as thick as $h_{\rm H_I}\sim3~{\rm kpc}$ and can have $h_{\rm H_I}\sim R$ in some regions. This implies that it may not be valid to approximate the H$_{\rm I}$ component to be a disc in the linear perturbation theory (see Section \ref{AppSingle}). However, the H$_{\rm I}$ gas has significantly lower surface densities than the other components, and therefore its contribution to the dynamical states of the arms is negligible (see below). The computed instability parameters indeed hardly change even if we assume a razor-thin arm for H$_{\rm I}$ by setting $h_{\rm H_I}=0$. Therefore, the possible invalidity of the disc approximation for the H$_{\rm I}$ component does not cause our analysis to deteriorate.

The bottommost panels of Fig. \ref{M51_Ingre} show the single-component instability parameters, $\min(S_i)$, where the blue colours indicate spiral-arm regions with $\min(S_i)<1$. Low values of $\min(S_i)<1$ predicting the instability are only seen in the small regions at the tips of the depicted spiral-arm regions of the H$_{\rm 2}$ component (the bottom left panel). It is worthy to mention that the value of $\min(S_{\rm H_2})$ of H$_{\rm 2}$ arms are significantly lower than those of the other components. The H$_{\rm 2}$ component indicates $\min(S_{\rm H_2})$ close to unity in most of the arm regions, whereas the H$_{\rm I}$ and stellar arms have $\min(S_{\rm H_I})\gtrsim10$ and $\min(S_{\rm s})\gtrsim3$. This means that the H$_{\rm I}$ and stellar components are highly stable and hardly contribute to decreasing the total instability parameters, $S_{\rm tot}$, of the arms (see equation \ref{S_total}). On the other hand, the molecular gas is the decisive component to determine the dynamical states of the arms. For the H$_{\rm I}$ gas, the high values of $\min(S_{\rm H_I})\gg1$ are primarily due to the significantly low $\Sigma_{\rm H_I}$. Although the stellar component has the highest $\Sigma_{\rm s}$ in the three components, its $\sigma_{\rm s}$ and $W_{\rm s}$ are also quite high. Hence, the values of $\min(S_{\rm s})$ of the stellar arms are high. Thus, the molecular gas traced by CO emission is the key component to determine the (in)stability of the inner arms of M51.

\begin{figure}
  \includegraphics[bb=0 0 2346 1430, width=\hsize]{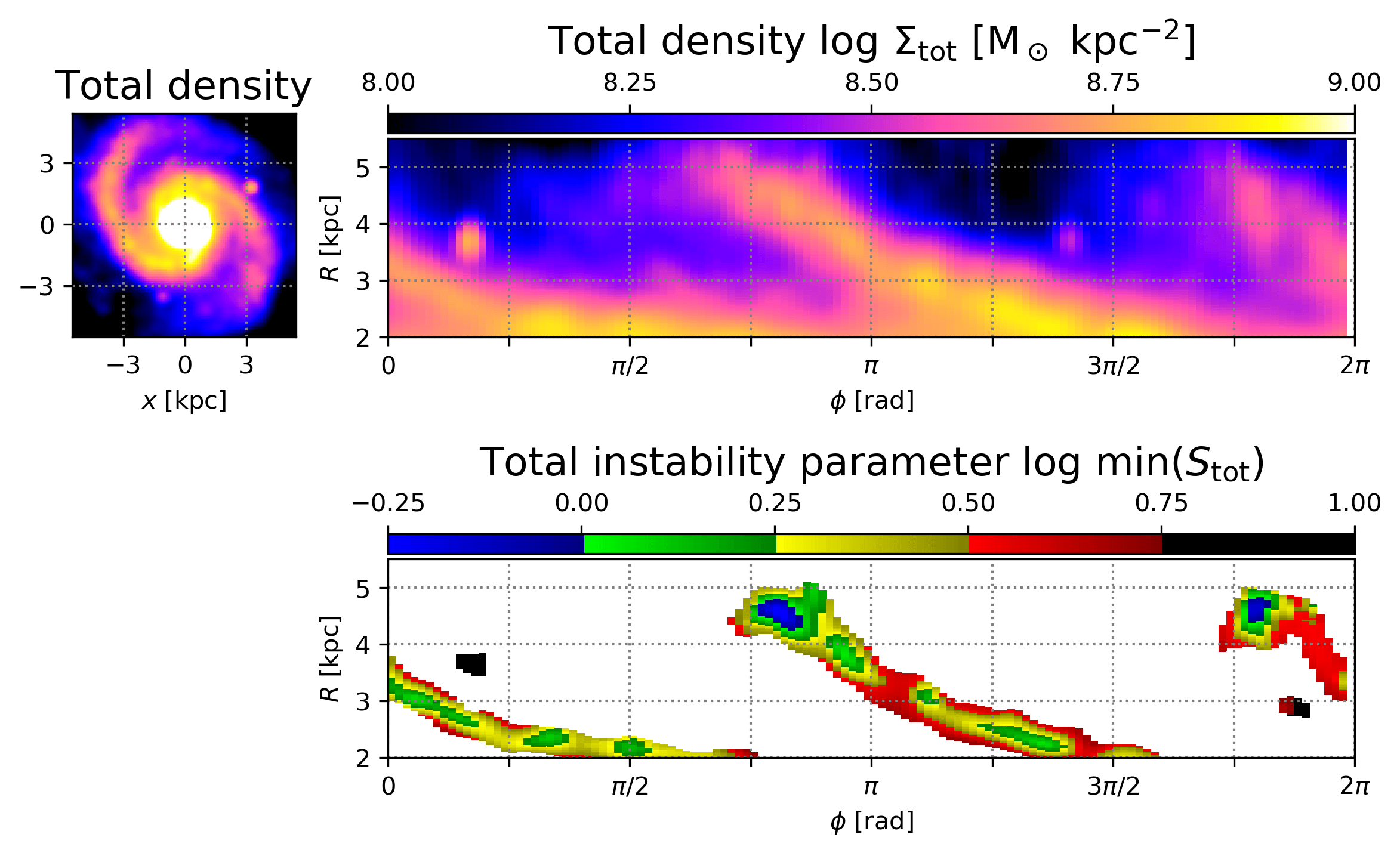}
  \caption{Distributions of the deprojected surface densities of the total baryon of M51 in the Cartesian (top left) and polar coordinates (top right). The bottom panel indicates the total instability parameters computed by equation (\ref{S_total}) in the spiral-arm regions; note that the colour scales are logarithmic. Only the spiral-arm regions are coloured, and the definition is the same as in Fig. \ref{M51_Ingre}.}
  \label{M51_SAI}
\end{figure}
Fig. \ref{M51_SAI} shows polar maps of the total surface densities $\Sigma_{\rm tot}\equiv\Sigma_{\rm H_2}+\Sigma_{\rm H_I}+\Sigma_{\rm s}$ and the total instability parameters $\min(S_{\rm tot})$ computed with equation (\ref{S_total}) in the spiral-arm regions. The values of $\min(S_{\rm tot})$ are only slightly lower than those of H$_{\rm 2}$ instability parameters $\min(S_{\rm H_2})$ in the bottom left panel of Fig. \ref{M51_Ingre}: $\min(S_{\rm tot})\simeq\min(S_{\rm H_2})$.

\begin{figure*}
  \includegraphics[bb=0 0 2067 1087, width=\hsize]{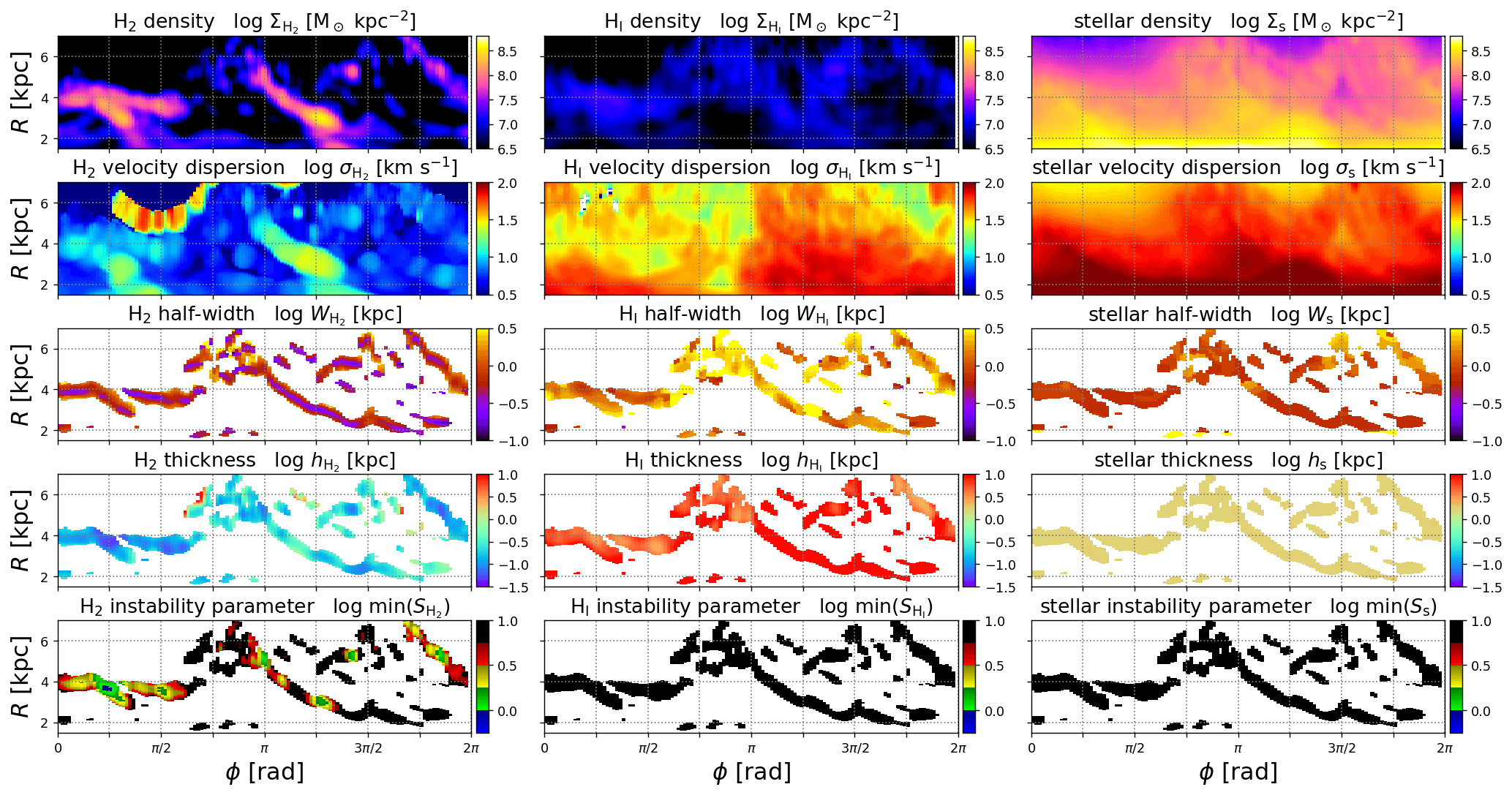}
  \caption{Same as Fig \ref{M51_Ingre} but for NGC3627. In the three rows from the bottom, the spiral-arm regions are defined by $\log[(\chi^2_{\rm H_2}+\chi^2_{\rm H_I})/2]>-0.25$. In the panel for $\sigma_{\rm H_2}$ (the second from the top in the left column), the H$_{\rm 2}$ velocity dispersions indicate quite high values in the regions around $(R,\phi)\sim(5$--$6~{\rm kpc},0.5\upi$--$0.7\upi)$ because of the influence by the edge of the observed field for CO(2-1) emission. Because no spiral arms are detected in these regions, our result is not affected by the edge effect.}
  \label{NGC3627_Ingre}
\end{figure*}
\begin{figure}
  \includegraphics[bb=0 0 2346 1430, width=\hsize]{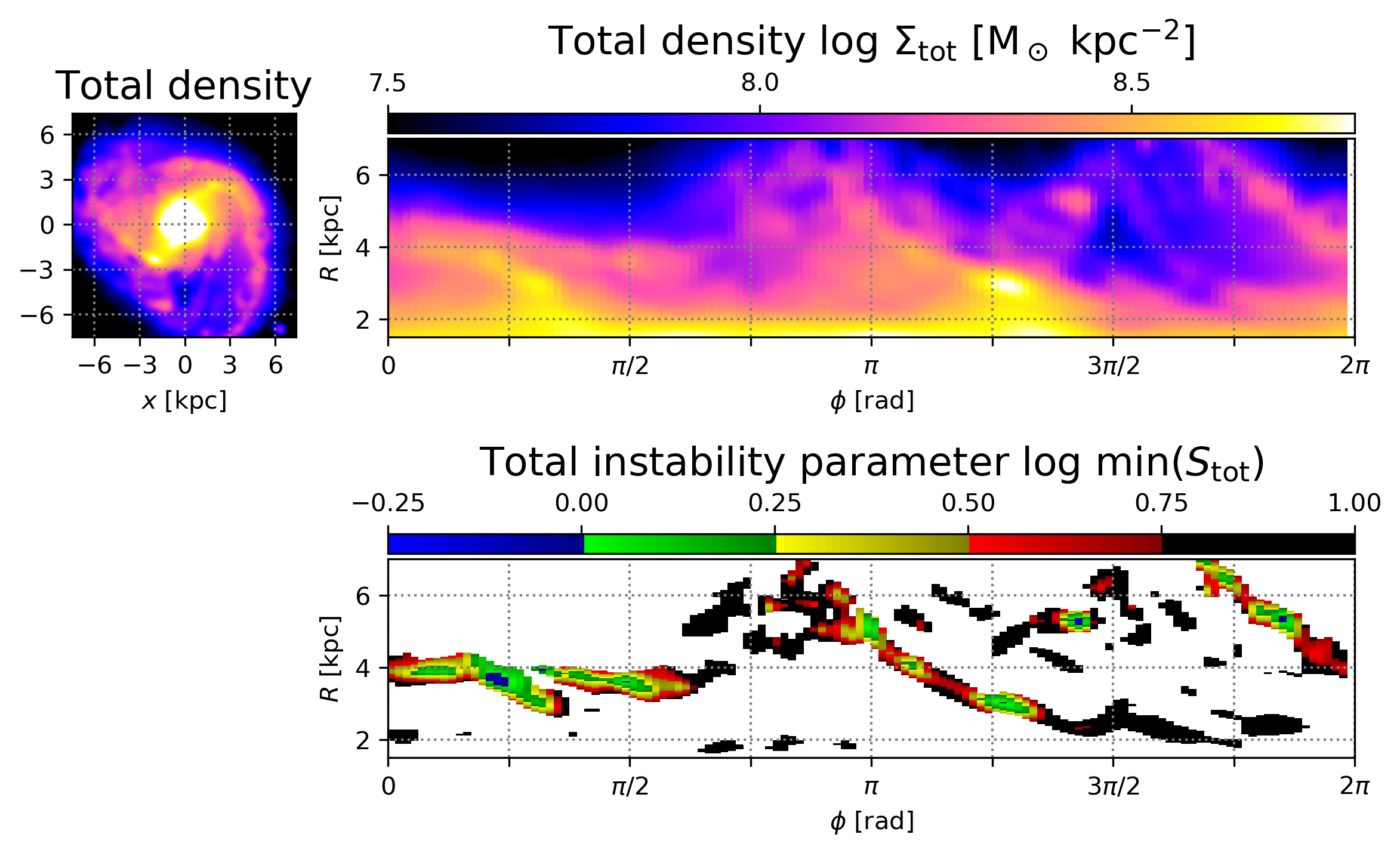}
  \caption{Same as Fig. \ref{M51_SAI} but for NGC3627.}
  \label{NGC3627_SAI}
\end{figure}

In the bottom panel of Fig. \ref{M51_SAI}, two areas indicate the instability with $\min(S_{\rm tot})<1$ at the tips of both inner arms: the blue areas at $(R,\phi)\simeq(4.5~{\rm kpc},0.8\upi~{\rm rad})$ and $(4.5~{\rm kpc},1.8\upi~{\rm rad})$.\footnote{We find the wavelengths of the most unstable perturbations in these regions to be $\lambda_{\rm MU}\sim1~{\rm kpc}$, which are comparable to the sizes of these areas with $\min(S_{\rm tot})<1$.} The low values of $\min(S_{\rm tot})$ appear to be due to the giant molecular clouds (GMCs) observed there (see the $\Sigma_{\rm H_2}$ map in the top left panel of Fig. \ref{M51_Ingre}), which are thought to be non-linear structures that have already collapsed. We cannot distinguish whether the GMCs are formed by SAI or other physical mechanisms. The other spiral-arm regions indicate $\min(S_{\rm tot})>1$ and are predicted to be dynamically stable. Thus, although our analysis is limited to the region inside $R\lesssim6~{\rm kpc}$ of the galaxy, we expect that the inner arms of M51 are not on the way to fragment from our SAI analysis.

We note, however, that our analysis in this paper involves various uncertainties of the observations and our modellings. Since the instability parameters, $\min(S_{\rm tot})$, in most of the arm regions are slightly above but close to unity indicating `marginally' stable states, the range of possible variation due to the uncertainties is thought to cover the boundary of $\min(S_{\rm tot})=1$. We discuss the possible uncertainties and the influence on $\min(S_{\rm tot})$ in Section \ref{uncertainties}. Besides, we do not argue the (in)stability of the outer arms in regions of $R\gtrsim6~{\rm kpc}$ in this study. We defer the SAI analysis for the outer arms of M51 to our future work.

\subsection{NGC3627}
\label{ngc3627res}
Fig. \ref{NGC3627_Ingre} shows our polar-map analysis for NGC3627. Similarly to M51, the H$_{\rm I}$ arms have quite low surface densities $\Sigma_{\rm H_I}$ although the H$_{\rm I}$ gas diffusely distributes along the molecular arms. The H$_{\rm I}$ velocity dispersions $\sigma_{\rm H_I}$ do not seem to show systematic differences between the arms and inter-arm regions.

\begin{figure*}
  \includegraphics[bb=0 0 2067 1087, width=\hsize]{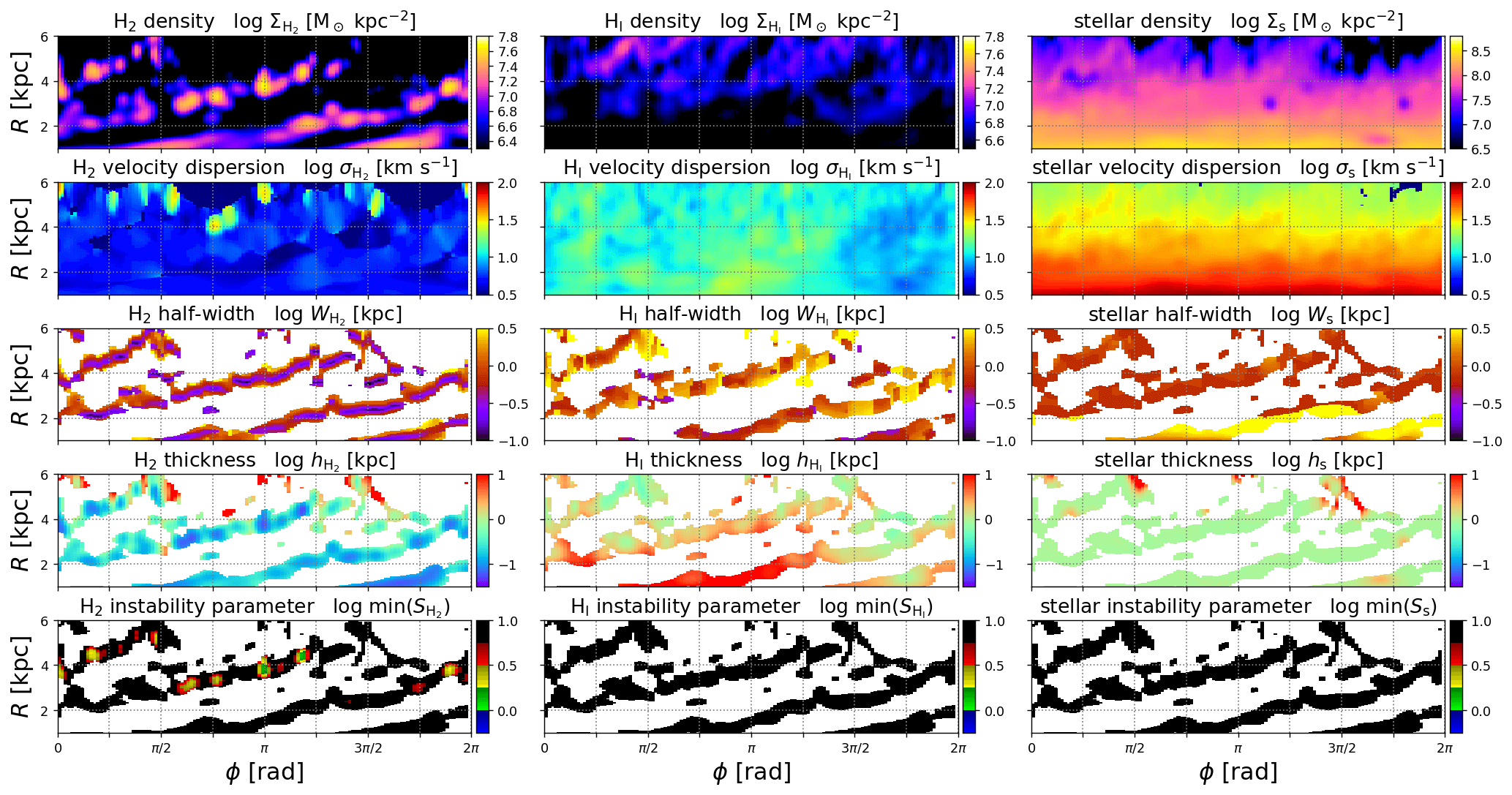}
  \caption{Same as Figs \ref{M51_Ingre} and \ref{NGC3627_Ingre} but for NGC628. In the three rows from the bottom, the spiral-arm regions are defined by $\log[(\chi^2_{\rm H_2}+\chi^2_{\rm H_I})/2]>-0.25$.}
  \label{NGC628_Ingre}
\end{figure*}

\label{ngc628res}
\begin{figure}
  \includegraphics[bb=0 0 2346 1430, width=\hsize]{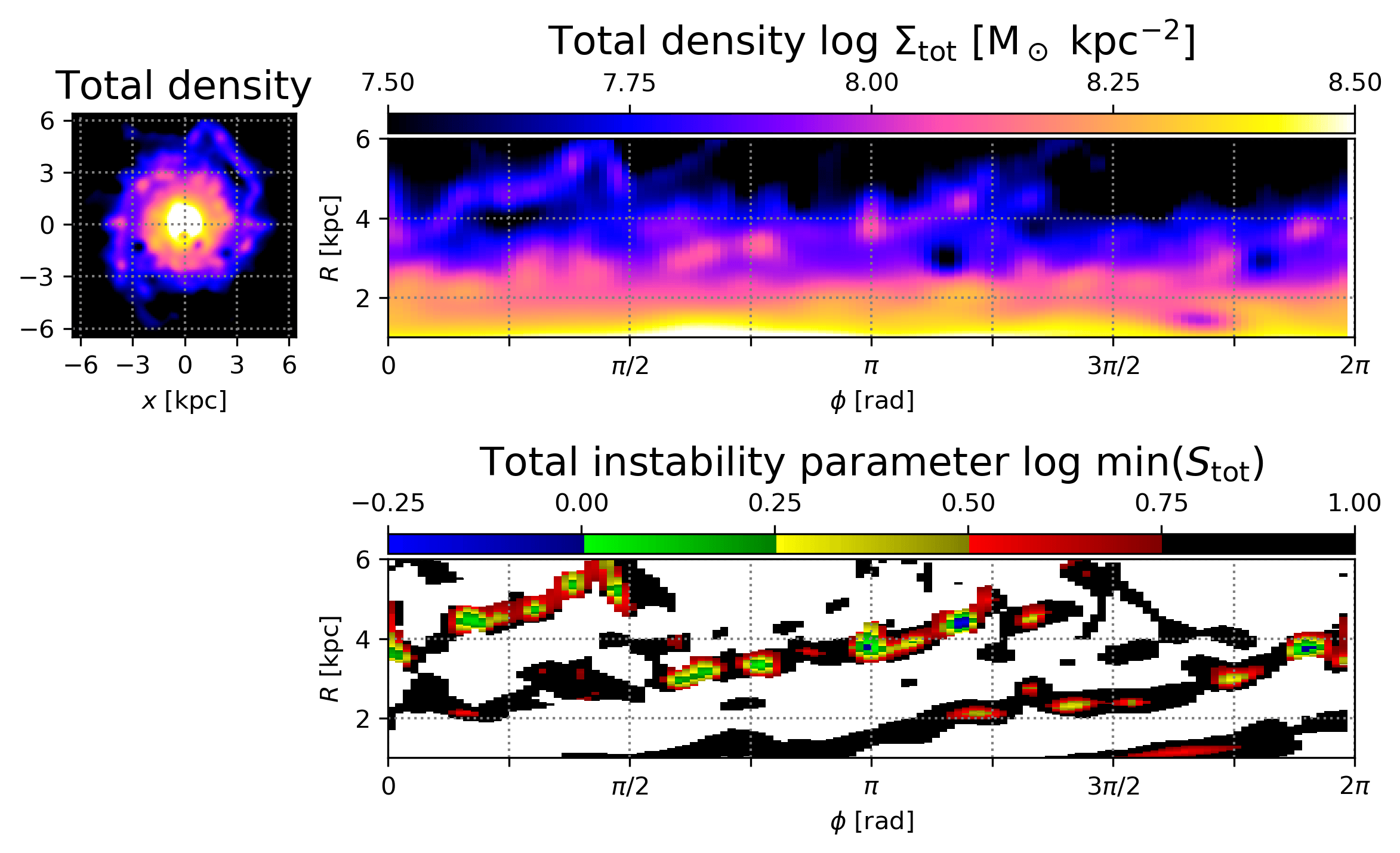}
  \caption{Same as Figs. \ref{M51_SAI} and \ref{NGC3627_SAI} but for NGC628.}
  \label{NGC628_SAI}
\end{figure}

The H$_{\rm 2}$ arms are thin and indicate low values of $W_{\rm H_2}$ in the third row in Fig. \ref{NGC3627_Ingre}. On the other hand, the stellar arms are hard to recognise because of the weak density contrast between the arms and the inter-arm regions. Therefore, our arm-detection scheme cannot capture the stellar arms well, and the goodness-of-fit $\chi^2_{\rm s}$ of the stellar component is generally higher than those of H$_{\rm 2}$ and H$_{\rm I}$. Accordingly, we impose a lower limit of $W_{\rm s}=0.5~{\rm kpc}$ to prevent the stellar arms from having erroneously small $W_{\rm s}$. We confirm that the lower limit hardly affects our results of $\min(S_{\rm tot})$. In the case of an extremely wide arm with $kW_i\gg1$, the SAI analysis does not break down but reduces to the Toomre instability analysis for azimuthal perturbations \citep[see Section \ref{AppSingle} and][]{iy:18}. Even if we manually set $W_{\rm s}$ to such a large value, we find that our result of $\min(S_{\rm tot})$ hardly changes since the stellar component is not dominant in the SAI analysis for this galaxy (see below). For this galaxy, we define the spiral arms to be the regions where $\log[(\chi^2_{\rm H_2}+\chi^2_{\rm H_I})/2]<-0.25$ without taking into account $\chi_{\rm s}$. The spiral-arm regions detected with this threshold correspond to the coloured regions in the three bottom rows of Fig. \ref{NGC3627_Ingre}. Our method appears to capture the two long spiral arms of NGC3627; it however also detects a number of small structures in the H$_{\rm 2}$ distribution, e.g. the regions around $(R,\phi)\sim(4$--$6~{\rm kpc},\upi$--$1.7\upi)$. These structures mostly seem to be GMCs and short filamentary structures in the inter-arm regions, and we do not argue the dynamical states of these small structures in this study.

As in the case of M51, the vertical thicknesses $h_{\rm H_I}$ of the H$_{\rm I}$ arms are quite large in Fig. \ref{NGC3627_Ingre}, and the disc approximation may not be appropriate for the H$_{\rm I}$ component. However, ignoring the thickness by setting $h_{\rm H_I}=0$ does not change the result of $\min(S_{\rm H_I})\gg1$. The H$_{\rm I}$ gas is thus negligible for the dynamical states (see below).

In the bottommost panels of Fig. \ref{NGC3627_Ingre}, both of the H$_{\rm I}$ and stellar components indicate the high values of $\min(S_i)\gg1$ in all of the spiral-arm regions. This means that the presence of these components little affects the (in)stability of the spiral arms in NGC3627. The molecular component, on the other hand, indicates $\min(S_{\rm H_2})$ close to unity in the spiral-arm regions. The H$_{\rm 2}$ gas is thus the crucial component for the dynamical states of the arms.

The bottom panel of Fig. \ref{NGC3627_SAI} shows the total instability parameters in the spiral-arm regions. Although most of the spiral-arm regions are in marginally stable states with $\min(S_{\rm tot})>1$, some small areas indicate $\min(S_{\rm tot})<1$, e.g. at $(R,\phi)\sim(4~{\rm kpc},0.25\upi)$. We estimate the wavelength of the most unstable perturbation to be $\lambda_{\rm MU}\sim1~{\rm kpc}$ in these areas. The size of these areas are smaller than the wavelength $\lambda_{\rm MU}$, and these areas are therefore actually expected to be stable. Moreover, these areas can indicate $\min(S_{\rm tot})>1$ if we take into account the uncertainties of the CO-to-H$_{\rm 2}$ conversion factor (Section \ref{alphaCO}). We thus expect that the arms of NGC3627 are marginally stable and not on the way to fragment.

\subsection{NGC628}
Fig. \ref{NGC628_Ingre} shows our polar-map analysis for NGC628. As we mentioned in Section \ref{ngc628}, this galaxy has a relatively lower surface density than the other two galaxies. Although the grand-design arms outshine in the H$_{\rm 2}$ distribution, the stellar arms are not clear in the map of $\Sigma_{\rm s}$. In inner radii of $R\lesssim2~{\rm kpc}$, the stellar arms are detected with erroneously large widths $W_{\rm s}\gtrsim3~{\rm kpc}$ (the third panel in the right column). Therefore, we define the spiral-arm regions in this galaxy by the threshold of $\log[(\chi^2_{\rm H_2}+\chi^2_{\rm H_I})/2]>-0.25$ without taking into account $\chi_{\rm s}$. The panels in the bottom three rows demonstrate that the threshold can capture the two spiral arms in this galaxy (the coloured regions) although some small-scale structures in the inter-arm regions are also detected.

The bottommost panels of Fig. \ref{NGC628_Ingre} show the single-component instability parameters, and those of the H$_{\rm I}$ and stellar components indicate $\min(S_i)\gg1$. This result implies that these components are negligible in the dynamical analysis for the arms. Although the H$_{\rm 2}$ arms indicate $\min(S_{\rm H_2})\sim1$--$3$ in some small areas, most of the spiral-arm regions have high values of $\min(S_{\rm H_2})$.

Fig. \ref{NGC628_SAI} shows the total instability parameters $\min(S_{\rm tot})$ for NGC628. The spiral-arm regions indicate high values of $\min(S_{\rm tot})>1$ predicting stable states for the SAI, except a few small areas that coincide with the positions of GMCs seen in the maps of $\Sigma_{\rm H_2}$ and $\Sigma_{\rm tot}$.\footnote{The GMCs are at $(R,\phi)\sim(4~{\rm kpc},\upi)$, $(4.5~{\rm kpc},1.2\upi)$ and $(4~{\rm kpc},1.8\upi)$} In the areas indicating $\min(S_{\rm tot})<1$, we find that the wavelengths of the most unstable perturbations are $\lambda_{\rm MU}\sim1~{\rm kpc}$, which are significantly larger than the sizes of the areas. Hence, these areas are not considered to be unstable. The grand-design arms of NGC628 are thus predicted to be stable. This result is similar to those of the other galaxies of M51 and NGC3627. Thus, none of the arms in our sample galaxies clearly indicates the instability with $\min(S_{\rm tot})<1$.

\section{Discussion}
\label{dis}
\subsection{Uncertainties and limitations of our SAI analysis}
\label{uncertainties}
\subsubsection{Uncertainties of the CO-to-H$_{\rm 2}$ conversion factor}
\label{alphaCO}
As we show in Section \ref{res}, molecular gas is the most important component in the SAI analysis for our sample galaxies, and atomic gas and stars do not largely contribute to the dynamical states of their arms. Uncertainties related to H$_{\rm 2}$ can therefore impact our results significantly. The most problematic uncertainties are those of the CO-to-H$_{\rm 2}$ conversion factor $\alpha_{\rm CO}$. Although we adopt our fiducial value of $\alpha_{\rm CO(1-0)}=3.5~{\rm M_\odot~pc^{-2}(K~km~s^{-1})^{-1}}$ measured in local regions along the spiral arms of M51 \citep{swa:10}. However, substantial variation of $\alpha_{\rm CO(1-0)}$ depending on measurements and environments has been reported in previous studies. Although \citet{bwl:13} have recommended $\alpha_{\rm CO(1-0)}=4.3~{\rm M_\odot~pc^{-2}(K~km~s^{-1})^{-1}}$ as a typical value for the Galactic inner disc, measurements by previous studies vary within $\pm0.3~{\rm dex}$ around this value. The median value among GMCs in nearby galaxies is $\alpha_{\rm CO(1-0)}=6.0~{\rm M_\odot~pc^{-2}(K~km~s^{-1})^{-1}}$ with a scatter of $0.4~{\rm dex}$ \citep{bwl:13}. In our computations, if we use a value twice as large as our fiducial $\alpha_{\rm CO(1-0)}$, considerable areas of the spiral-arm regions of M51 and NGC3627 indicate the instability with $\min(S_{\rm tot})<1$. Hence, if we take the uncertainties and/or variation of $\alpha_{\rm CO(1-0)}$ into consideration, we cannot necessarily reject the possibility that the spiral arms of M51 and NGC3627 are dynamically unstable. 

\begin{figure}
  \includegraphics[bb=0 0 2105 1555, width=\hsize]{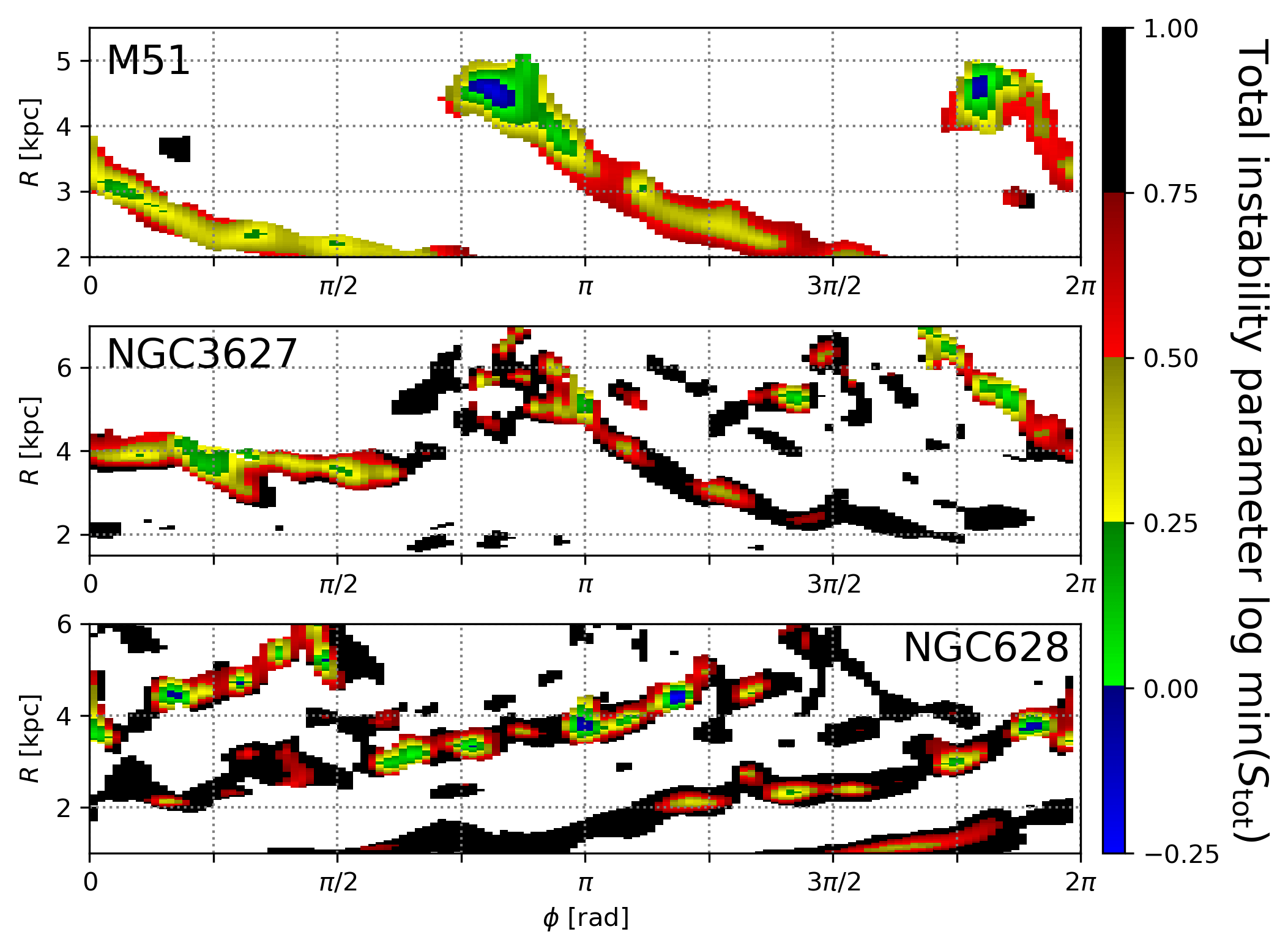}
  \caption{Same as the bottom panels of Figs. \ref{M51_SAI}, \ref{NGC3627_SAI} and \ref{NGC628_SAI} for the total instability parameters but computed with the variable CO-to-H$_{\rm 2}$ conversion factor given by equation (\ref{Narayanan2}).}
  \label{VariableAlpha}
\end{figure}
Although we assume the uniform value for $\alpha_{\rm CO}$ in this study, the conversion factor can depend on environment and physical condition, generally decreases with increasing gas density and metallicity \citep[e.g.][]{nko:12,bwl:13,lnd:18,iyy:20}. Dense gas in spiral arms may therefore be expected to have a relatively lower $\alpha_{\rm CO}$ than in inter-arm regions. Using a large set of galaxy simulations and radiation transfer computations, \citet{nko:12} have proposed variable $\alpha_{\rm CO(1-0)}$ as a function of brightness temperature of CO line and metallicity,
\begin{equation}
\alpha_{\rm CO(1-0)} = \frac{\min\left(6.3,10.7\times W_{\rm CO}^{-0.32}\right)}{\left( Z/Z_\odot\right)^{0.65}},
\label{Narayanan2}
\end{equation}
where $W_{\rm CO}$ is line intensity of CO(1-0) in units of ${\rm K~km~s^{-1}}$ and $Z/Z_\odot$ is gas-phase metallicity with respect to the solar value.\footnote{\citet{iyy:20} postprocessed data of a cosmological simulation and found that the galaxy-integrated values of $\alpha_{\rm CO(1-0)}$ also approximately follow equation (\ref{Narayanan2}) for massive  galaxies at redshift $z=0$.} To evaluate the influence of the local variation of the conversion factor, we adopt equation (\ref{Narayanan2}) to our zeroth-moment maps of CO lines while ignoring the metallicity-dependence with $Z/Z_\odot=1$. Fig. \ref{VariableAlpha} shows the results of the total instability parameters using equation (\ref{Narayanan2}) for $\Sigma_{\rm H_2}$ in the three galaxies; we again assume $r_{\rm 21}=0.79$ for NGC3627. The results are hardly different from those in the fiducial cases in Figs. \ref{M51_SAI}, \ref{NGC3627_SAI} and \ref{NGC628_SAI}. Most of the spiral-arm regions indicate high values of $\min(S_{\rm tot})>1$ predicting their stable states except inside a few GMCs. However, note that we ignore the metallicity-dependence in equation (\ref{Narayanan2}).

In addition to $\alpha_{\rm CO(1-0)}$, for NGC3627, we assume the CO(2-1)-to-CO(1-0) line ratio $r_{\rm 21}=0.79$ measured from the unbiased galaxy sample of the xCOLDGASS survey \citep{xclodgass:17}. Analysis of \cite{lws:13} based on the HERACLES survey also proposes a similar value of $r_{\rm 21}=0.7$. Localised simulations of \citet{gokk:20}, however, showed significant variation of $r_{\rm 21}$ depending on various physical parameters. Besides, \citet{bcb:21} observationally obtained two-dimensional distributions of $r_{\rm 21}$ in nine local galaxies and estimated luminosity-weighted averages for the individual galaxies range from $r_{\rm 21}=0.48$ to $0.73$. They also showed radial variation of $r_{\rm 21}$ in some galaxies and estimated a typical azimuthal variation to be $\sim20$ per cent. Thus, the dynamical analysis using CO(2-1) observations, such as NGC3627 in this study, is involved with another uncertainty of the line ratio $r_{\rm 21}$. Furthermore, it should be noted that the second moments, $\sigma_{\rm H_2}$, can be systematically different between measurements using CO(1-0) and CO(2-1) lines since CO(2-1) traces denser gas than CO(1-0).

\subsubsection{Other uncertainties and limitations}
\label{others}
Distance measurements directly affect the determinations of length-scales in galaxies. Although surface density $\Sigma_i$ is basically independent of distance $d$, arm width $W_i$ increases with $d$. Since some physical quantities in equation (\ref{S_each}) vary with $W_i$, such as line-mass $\Upsilon_i\propto\Sigma_iW_i$, uncertainties of $d$ can affect the instability parameter $S_{\rm tot}$. Even though all of our sample galaxies are nearby spirals at $d\lesssim10~{\rm kpc}$, their distances are still somewhat imprecise. For example, we assume $d=7.6~{\rm Mpc}$ for M51 following the PAWS observations \citep{cfj:02}, whereas the PHANGS project adopts $d=8.56~{\rm Mpc}$ \citep{alv:20}, and \citet{mwf:12} assume $d=9.9~{\rm Mpc}$ \citep{tgt:09}. We refer the readers to figure 3 of \citet{msd:16} for previous distance measurements of M51. If we assume $d=9.9~{\rm Mpc}$ for M51, our result indicates $\min(S_{\rm tot})<1$ in wide regions of the inner arms; however the impact of such a distance uncertainty is less significant than that of $\alpha_{\rm CO}$ argued in Section \ref{alphaCO}. For NGC3627, the scatter of distance measurements in literature seems to be relatively small. \citet{alv:20} have derived $d=11.07\pm0.44~{\rm Mpc}$ using tip of red giant branch method, which is consistent with $d=10.6~{\rm Mpc}$ adopted in this work. Other measurements using different methods appear to converge at $d\simeq10$--$11~{\rm Mpc}$ \citep[e.g.][]{sst:99,fmg:01,fmh:19,hfm:19}. Thus, for NGC3627, the distance uncertainty does not change our result. For NGC628, we refer the readers to table 4 of \citet{kgb:17} for the distance measured with various methods, where the previous determinations range from $d=7.3$ to $10.2~{\rm Mpc}$. Recent observations appear to prefer relatively large distances of $d\simeq9$--$10~{\rm Mpc}$, which are close to $d=9.8~{\rm Mpc}$ in this study. In our analysis, adopting a lower $d$ results in a higher $\min(S_{\rm tot})$. Hence, even if we assume $d=7.31~{\rm Mpc}$ of \citet{skt:96}, our analysis does not predict the instability for NGC628.

Although our analysis posits the tight-winding approximation, it is not necessarily justified for the grand-design arms in this study. However, \citet{iy:18} have demonstrated that the SAI analysis can characterise the instability with accuracy for arms with relatively large pitch angles in their simulations. It is probably because strengths of the relevant forces in the analysis, i.e. gravity, pressure and Coriolis force, are basically independent of directions, as long as $\sigma_{\rm s}\simeq\sigma_{R,{\rm s}}$. The dispersion relation (equation \ref{omega_eachl} in Appendix \ref{AppMulti}) derived from the linear perturbation theory is indeed invariant even if the perturbations are radial. However, the arm widths $W_i$ depend on pitch angle $\theta$ since these are measured with the radial fitting for $\Sigma_i$ (see Section \ref{armfit}). The half-widths $W_i$ are overestimated by a factor of $1/\cos{\theta}$, which underestimates $S_i$.

In Section \ref{armfit}, our analysis implicitly assumes that density distribution of a spiral arm is symmetric between its inner and outer sides (i.e. the radial Gaussian distribution). \citet{smm:20} measured widths of stellar spiral arms of local galaxies and found that outer sides generally have larger $W_{\rm s}$ than their inner sides \citep[see also][]{msm:20} although the differences are only $14$ per cent on average. It is unclear whether the symmetric assumption is appropriate for the gas components. Our analysis may have room for improvement to take into account such asymmetry of a spiral arm. 

\citet{emk:17} found that one of the inner arms of M51 has an offset of density peaks between its gas and stellar components. Although the observed offset is smaller than the widths of the spiral-arm regions in Figs \ref{M51_Ingre} and \ref{M51_SAI}, our analysis cannot treat offsets of spiral arms between different components and therefore assumes that there is no offset. We show, however, that the stellar components are far less important than H$_{\rm 2}$ in the galaxies. We therefore expect that such an offset between stellar and gas components little affects our results.

In Section \ref{kapa}, we assume rigid rotations within the spiral arms: $\kappa=2\Omega$ for equation (\ref{S_each}). Although this assumption is motivated by the results of isolated simulations of \citet{tti:16} and \citet{iy:18}, it may not be appropriate for the sample galaxies in this study. Accordingly, we recompute the instability parameters without this assumption but with equation (\ref{kappa}) for $\kappa$. We find that $\min(S_{\rm tot})$ becomes slightly lower but does not change significantly in the sample galaxies.

We ignore the contributions of thermal and magnetic pressure to velocity dispersions of the gas components (Section \ref{disp}). Including thermal (magnetic) pressure can significantly increase $\min(S_{\rm tot})$ if its sound (Alfv\'{e}n) velocity is comparable to the kinetic velocity dispersion. Meanwhile, effects of magnetic fields are complicated; toroidal fields can destabilise spiral arms by cancelling Coriolis force \citep{e:87,e:94,iy:18b}. However, molecular gas is generally cold therefore does not have a such high sound velocity. Since magnetic fields are generally observed to be quite weak in local galaxies \citep[e.g.][]{h:17}, the magnetic pressure and destabilisation are thought to be dynamically negligible \citep[e.g.][]{kk:18}.  

We use the stellar disc model of \cite{lwb:08} to estimate azimuthal and vertical velocity dispersions of the stellar components (Section \ref{disp}). We consider that the uncertainties of the model would be quite large, which can involve various systematic errors. Moreover, it is not justified whether we can apply the `disc' model to the spiral arms. However, as we mention above and in Section \ref{res}, stellar components are of secondary importance in our SAI analysis for the sample galaxies. Therefore, the impact of the model uncertainties would be limited and smaller than that of $\alpha_{\rm CO}$. Observations for stars with integral field units enable us to measure two-dimensional distribution of line-of-sight velocity dispersions. Although we still need to assume velocity ellipsoids ($\sigma_{z,{\rm s}}/\sigma_{\rm s}$ and $\sigma_{R,{\rm s}}/\sigma_{\rm s}$), these observations are expected to significantly improve our computations for $S_{\rm s}$.

The local analysis of SAI is based on the local properties of a spiral arm. This ensures the flexibility of our analysis with respect to the global properties of galaxies. The accuracy of the analysis using equations (\ref{S_each} and \ref{S_total}) has been confirmed in numerical simulations \citep{iy:18,iy:20}, and \citet{iy:18} applied it to the galactic models motivated by high-redshift discs and the formation of giant clumps. Although such high-redshift discs can be different from the local galaxies in various properties, \citet{iy:18} demonstrated that the analysis is still accurate even in pure stellar discs and those with largely different rotation curves (bulge-to-disc mass ratios). A possible problem is that we applied the azimuthally averaged rotation velocities to each radius in this study. The companion galaxies that drove the formation of the grand-design arms can disturb the local velocity fields of the galaxies. Note that our analysis in this paper has missed this effect. For M51, \citet{os:14} have argued that disturbance by the companion is seen in the outer radii $R\gtrsim7.5~{\rm kpc}$, and therefore we did not apply the analysis to the radii. However, the other galaxies lack such detailed analysis for local velocities.

\subsection{Are spiral arms of local galaxies stable?}
\label{stability}
This study aims to answer this question by means of the SAI analysis for the local galaxies. Our sample is, however, comprised of only a handful of spiral arms in the three grand-design spiral galaxies. We therefore cannot state a general answer to the question for local spiral galaxies in this paper. In addition, this study does not argue the other types of spiral galaxies such as multi-arm ($m\geq3$)\footnote{Here, $m$ represents the number of arms in a galaxy.} and flocculent spirals. The formation of these spiral arms is probably driven by internal dynamics of galactic discs, whereas that of grand-design arms is thought to be external interactions with other galaxies \citep[e.g.][]{dobbsbaba:14}. It is therefore intriguing to apply the SAI analysis to the other types of spirals in our future studies.

Although we need to be careful of the uncertainties of $\alpha_{\rm CO}$, none of the spiral arms indicates significant instability with $\min(S_{\rm tot})<1$ in this study, except inside the two GMCs of M51 (Section \ref{m51res}). This is, so far, consistent with our expectation in Section \ref{Intro} and the rarity of the disc galaxies hosting giant clumps of $M_{\rm clump}\gtrsim10^8{\rm M_\odot}$ in the local Universe \citep[e.g.][]{sok:16,f:17,fga:17}. \citet{dobbsbaba:14} have argued that lifetimes of grand-design spiral arms would be as long as $\sim1~{\rm Gyr}$ since they are thought to form by encounters of companion galaxies. On the other hand, a spiral arm fragments in nearly local dynamical time scales, $\sim100~{\rm Myr}$, after the arm turns into an unstable state with $\min(S_{\rm tot})<1$ \citep{iy:18}. Therefore, if grand-design arms are generally unstable and fragment in such a short time scale, the prevalence of grand-design spiral galaxies in the local Universe could not be explained. In this sense, our result is consistent with this picture.  

Our results show that molecular gas is the decisive component for the dynamical states of the arms in our sample galaxies. The total instability parameters are only slightly lower than those of their H$_{\rm 2}$ components, i.e. $\min(S_{\rm tot})\simeq\min(S_{\rm H_2})$. This implies that atomic gas and stellar components are not important and could be ignored. In applying the SAI analysis to an observed galaxy, the most painstaking and time-consuming tasks are to gather multi-band photometric data and to perform the two-dimensional SED fitting to obtain a stellar density map. If we could omit these procedures, we would be able to efficiently apply the analysis to a larger number of spiral galaxies. Recently, enormous datasets of CO line imaging of a number of nearby galaxies were made publicly available, such as COMING \citep[CO multi-line imaging of nearby galaxies, ][]{COMING} and PHANGS-ALMA. Utilising such data may allow us to statistically argue the (in)stability of spiral arms in local galaxies, which can lead us to `the general answer' to the aforementioned question. In addition, ALMA can resolve spiral arms in galaxies at intermediate redshifts of $z\lesssim1$. It may enable us to study the redshift-evolution of the dynamical states of spiral arms and solve the puzzle of the missing link between the high-redshift clumpy and low-redshift spiral galaxies.

\section{Conclusions and Summary}
\label{con}
Based on the multi-component linear perturbation analysis of \citet{iy:18,iy:20}, we investigate the dynamical states of the spiral arms in nearby galaxies: M51 (the inner arms), NGC3627 and NGC628. They are grand-design spiral galaxies whose arms were possibly formed by interactions with other galaxies. We take into account the three components of molecular gas, atomic gas and stars in the analysis, utilise their archived data of CO emission, 21-centimetre lines and multi-band photometric observations. 

Our analysis predicts the dynamical instability for a local region in a spiral arm when the instability parameter $\min(S_{\rm tot})<1$ and vice versa. From the analysis, our results show that the spiral arms of the three galaxies indicate marginally stable states with $\min(S_{\rm tot})>1$. Only the small regions inside the two GMCs in the inner arms of M51 indicate the instability with $\min(S_{\rm tot})<1$; such GMCs are, however, considered to be non-linear structures that have already collapsed. Thus, the spiral arms in the galaxies are predicted to be stable therefore not on the way to fragment. The stability predicted by our analysis for the galaxies is consistent with the rarity of clumpy galaxies and the prevalence of grand-design spirals in the local Universe.

For all of the spiral arms in our sample, our analysis shows that molecular gas is the decisive component for their dynamical states, whereas atomic gas and stellar components do not significantly contribute to the (in)stability of the arms. The importance of molecular gas implies that the accuracy of a CO-to-H$_{\rm 2}$ conversion factor affects our results significantly. Although our analysis indicates the marginally stable states with $\min(S_{\rm tot})$ slightly higher than unity, it can be turned into the prediction of unstable states with $\min(S_{\rm tot})<1$ if we take the uncertainties of the conversion factor into consideration. Hence, although the results adopting our fiducial value of $\alpha_{\rm CO(1-0)}=3.5~{\rm M_\odot~pc^{-2}(K~km~s^{-1})^{-1}}$ predict the stability, more robust dynamical analysis for spiral arms requires more accurate determinations of the conversion factor. 

\section*{Acknowledgements}
We are grateful to an anonymous reviewer for his/her efforts to improve this paper. We thank David V. Stark, Yusuke Miyamoto, Ken-ichi Tadaki and Naoki Yoshida for their valuable discussion. This study was supported by National Astronomical Observatory of Japan (NAOJ) ALMA Scientific Research Grant Number 2019-11A. SI receives the funding from KAKENHI Grant-in-Aid for Young Scientists (B), No. 17K17677, FE is financially supported by the Japan Society for the Promotion of Science (JSPS) KAKENHI Grant Number 17K14259, and HY receives the funding from Grant-in-Aid for Scientific Research (No. 17H04827 and 20H04724) from JSPS. This paper used the following ALMA data: ADS/JAO.ALMA \#2015.1.00956.S (PI: A. K. Leroy) and \#2013.1.00532.S (PI: E. Schinnerer). ALMA is a partnership of ESO (representing its member states), NSF (USA) and NINS (Japan), together with NRC (Canada), MOST and ASIAA (Taiwan), and KASI (Republic of Korea), in cooperation with the Republic of Chile. The Joint ALMA Observatory is operated by ESO, AUI/NRAO and NAOJ. Data analysis was carried out on the Multi-wavelength Data Analysis System operated by the Astronomy Data Center (ADC), NAOJ.

%%%%%%%%%%%%%%%%%%%%%%%%%%%%%%%%%%%%%%%%%%%%%%%%%%
\section*{Data availability}
The data underlying this article will be shared on reasonable request to the corresponding author.



\bibliographystyle{mnras}
%\bibliography{references}

%\begin{thebibliography}{99}
%\end{thebibliography}

%%%%%%%%%%%%%%%%%%%%%%%%%%%%%%%%%%%%%%%%%%%%%%%%%%

%%%%%%%%%%%%%%%%% APPENDICES %%%%%%%%%%%%%%%%%%%%%
%
\appendix

\section{Linear perturbation analysis}
\label{lpa}

In the SAI analysis, a spiral arm is assumed to have a negligibly small pitch angle (i.e. the tight-winding approximation), and materials in the arm are rotating on a disc plane with an azimuthal velocity $v_\phi$. With this assumption, the arm is approximated as a structure resembling a ring. In the polar coordinates $(R,\phi)$, the analysis considers azimuthal perturbations propagating along the arm, and their amplitudes are assumed to be proportional to $\exp[\mathrm{i}(ky-\omega t)]$, where $y\equiv\phi R$, and $\omega$ and $k$ are angular frequency and wavenumber of the azimuthal perturbation. If the wavelength $2\pi/k$ is shorter than the radius of the arm, i.e. $kR\gg1$, the curvature of the spiral arm is negligible. 

In what follows, we perform linear perturbation analysis to obtain the dispersion relation which describes $\omega$ as a function of $k$. If $\omega$ has a solution with a negative imaginary part at $k$, the amplitude of the perturbation $k$ is expected to grow exponentially with time, i.e. dynamical instability. 

\subsection{Single-component analysis}
\label{AppSingle}
By considering a single-component spiral arm of barotropic fluid with the assumptions described above, the linearised equations of continuity and radial and azimuthal momenta are given as 
\begin{equation}
\omega\delta \Upsilon = k\Upsilon\delta v_\phi,
\label{linearlized1}
\end{equation}
\begin{equation}
-\mathrm{i}\omega\delta v_R = 2\Omega\delta v_\phi
\label{linearlized2}
\end{equation}
and
\begin{equation}
-\mathrm{i}\omega\delta v_\phi = -2\Omega\delta v_R - \mathrm{i}k\frac{\sigma^2}{\Upsilon}\delta \Upsilon - \mathrm{i}k\delta\Phi,
\label{phimom}
\end{equation}
respectively (see \citealt{tti:16} and \citealt{iy:18} for the details), where the prefix $\delta$ means the perturbation of a physical value following it. For the perturbed potential $\delta\Phi$, the Poisson equation for a `ring' with a Gaussian density distribution and a vertical thickness $h$ is given as
\begin{equation}
\label{poisson}
  \delta\Phi =-\upi G\delta\Upsilon\left[K_0(kW)L_{-1}(kW) + K_1(kW)L_0(kW)\right]F(kh),
\end{equation}
where $K_j$ and $L_j$ are modified Bessel and Struve functions of order $j$ (see \citealt{tti:16} for the derivation of equation \ref{poisson}); hereafter we abbreviate $f(kW)\equiv K_0(kW)L_{-1}(kW) + K_1(kW)L_0(kW)$. The thickness correction factor $F(kh)=[1-\exp(-kh)]/(kh)$ \citep[see][]{t:64}. Combining equations (\ref{linearlized1} -- \ref{poisson}), the dispersion relation of the azimuthal perturbation $k$ is derived as 
\begin{equation}
\omega^2 = \left[\sigma^2-\upi G\Upsilon f(kW)F(kh)\right]k^2 + \kappa^2,
\label{DR1}
\end{equation}
If $\omega^2<0$, $\omega$ has a solution of a negative imaginary number, and the perturbation $k$ is unstable. Equation (\ref{DR1}) is transformed as
\begin{equation}
\frac{\sigma^2k^2 + \kappa^2 - \omega^2}{\upi G\Upsilon f(kW)F(kh) k^2}=1.
\label{DL11}
\end{equation}
The perturbation $k$ always satisfies this relation. In the boundary case of $\omega^2=0$, the above becomes
\begin{equation}
S(k)\equiv\frac{\sigma^2k^2 + \kappa^2}{\upi G\Upsilon f(kW)F(kh) k^2}=1.
    \label{S_eachl}
\end{equation}
By comparing with equation (\ref{DL11}), one can see that $S<1$ when $\omega^2<0$. Conversely, $S>1$ when $\omega^2>0$. Thus, one can consider $S$ as the instability parameter of SAI. Because equation (\ref{S_eachl}) is always a downward-convex function of $k$, $\min[S(k)]$ exists and can be used as the single-component instability parameter.

The function $f(kW)$ decreases monotonically with $kW$ and approaches asymptotically to $(kW)^{-1}$ when $kW\gg1$ \citep[see figure 1 of][]{iy:18}. In this case, a perturbation $k$ is deeply embedded within an arm width $W$, and the dispersion relation becomes independent of $W$. Namely, the SAI analysis reduces to Toomre instability analysis for azimuthal perturbations.

\subsection{Multi-component analysis}
\label{AppMulti}
We extend the above instability analysis to a multi-component system and give the $i$-th component its own physical properties: $\Upsilon_i$, $\sigma_i$, $\kappa_i$, $W_i$ and $h_i$. We consider that they interact with each other only through gravity and are connected via the Poisson equation. The perturbation of the total potential is the summation of those of all components,
\begin{equation}
    \delta\Phi_{\rm tot} = \sum_i\delta\Phi_i = -\sum_i \upi G\delta\Upsilon_if(kW_i)F(kh_i).
    \label{multiPoisson}
\end{equation}
In the multi-component system, $\delta\Phi$ in equation (\ref{phimom}) is replaced with $\delta\Phi_{\rm tot}$. Combining equations (\ref{linearlized1}, \ref{linearlized2} and \ref{phimom}) for the $i$-th component, we obtain
\begin{equation}
    \delta \Upsilon_i = \frac{k^2\Upsilon_i}{\omega^2-\kappa_i^2-\sigma_i^2k^2}\delta\Phi_{\rm tot}.
    \label{omega_eachl}
\end{equation}
Substituting equation (\ref{omega_eachl}) into equation (\ref{multiPoisson}), we obtain
\begin{equation}
    \sum_i\frac{\upi G\Upsilon_if(kW_i)F(kh_i)k^2}{\kappa_i^2+\sigma_i^2k^2 - \omega^2} = 1.
    \label{multiDR}
\end{equation}
This corresponds to equation (\ref{DL11}) in the single-component analysis. In the boundary case of $\omega^2=0$, we can define the instability parameter in the multi-component analysis as
\begin{equation}
    S_{\rm tot}(k) = \left[
    \sum_i\frac{1}{S_i(k)}\right]^{-1}.
    \label{sumS}
\end{equation}
This corresponds to equation (\ref{S_total}) in Section \ref{ana}. The perturbation $k$ is unstable when $S_{\rm tot}(k)<1$.

\citet{iyh:21} applied the SAI analysis to their simulations of collisional ring galaxies and show it to successfully characterise the fragmentation of their ring structures. \cite{iy:19} developed the analysis to include effects of toroidal magnetic fields which can destabilise spiral arms. They confirmed the accuracy of the magnetic analysis with ideal magnetohydrodynamics simulations of disc galaxies. Although the analysis in this study does not consider the magnetic effect, we discuss the influence in Section \ref{others}.

%If you want to present additional material which would interrupt the flow of the main paper, it can be placed in an Appendix which appears after the list of references.

%%%%%%%%%%%%%%%%%%%%%%%%%%%%%%%%%%%%%%%%%%%%%%%%%%

% Don't change these lines
\bsp	% typesetting comment
\label{lastpage}
\end{document}